\documentclass[ superscriptaddress, reprint, amsmath,amssymb, aps,pra,floatfix,]{revtex4-1}
\pdfoutput=1
\usepackage{graphicx}
\usepackage{float}
\usepackage{wrapfig}
\usepackage{subcaption}
\usepackage{dcolumn}
\usepackage{bm}
\usepackage{hyperref}
\usepackage{natbib}
\usepackage{physics}
\usepackage{amsmath}
\begin{document}
\title{Optimal control of interacting quantum systems based on the first-order Magnus approximation: Application to multiple dipole-dipole coupled molecular rotors}

\author{Andrew Ma}
\email{andrew.ma@princeton.edu}
\affiliation{%
Department of Physics, Princeton University, Princeton, New Jersey 08544, USA\\
}%
\author{Alicia B. Magann}
\email{amagann@princeton.edu}
\affiliation{%
Department of Chemical and Biological Engineering, Princeton University, Princeton, New Jersey 08544, USA\\
}%
\author{Tak-San Ho}
\email{tsho@princeton.edu}
\author{Herschel Rabitz}
\email{hrabitz@princeton.edu}
\affiliation{%
Department of Chemistry, Princeton University, Princeton, New Jersey 08544, USA\\
}%

\date{\today}
\begin{abstract}
We develop a methodology for performing approximate optimal control simulations for quantum systems with multiple interacting degrees of freedom.  The quantum dynamics are modeled using the first-order Magnus approximation in the interaction picture, where the interactions between different degrees of freedom are treated as the perturbation.  We present a numerical procedure for implementing this approximation, which leverages the separability of the zeroth-order time evolution operator and the pairwise nature of common interactions for a reduced computational cost.  This formulation of the first-order Magnus approximation is suitable to be combined with gradient-free methods for control field optimization; to this end, we adopt a Stochastic Hill Climbing algorithm.  The associated computational costs are analyzed and compared with those of the exact simulation in the large $N$ limit.  For numerical illustrations, we perform approximate optimal control simulations for systems of two and three dipole-dipole coupled molecular rotors under the influence of a global control field.  For the two rotor system, we optimize fields for both orientation and entanglement control objectives.  For the three rotor system, we optimize fields for orienting rotors in the same direction as well as in opposite directions.
\end{abstract}
\maketitle

\section{Introduction \label{Section_Introduction}}

There is considerable interest in using optimally shaped fields to control the dynamics of quantum systems \cite{Brif,Kosloff,Peirce,Shi,Werschnik}.  Such fields can be found using iterative optimization methods, including gradient~\cite{Khaneja,Maday,Rothman,Zhu} and gradient-free methods \cite{Caneva,Krieger,Magann}, and have potential applications in a wide range of areas, such as quantum information processing \cite{Arenz,Dolde,Waldherr}, chemical reactions \cite{Tannor}, semiconductor electron states~\cite{Rasanen}, nanostructured materials~\cite{Grigorenko}, and high-harmonic generation \cite{Winterfeldt}.

From a theoretical perspective, the study of controlling quantum systems with interacting degrees of freedom is particularly challenging, since the dimensionality of the state space scales exponentially with the number of degrees of freedom \cite{Brif,Caneva_2,Doria}.  This fact can  render exact optimal control simulations for many-body quantum systems computationally intractable.  To address this issue, a variety of approximate methods have been employed, including classical and semiclassical Gaussian wave-packet approximations \cite{Kohler,Messina}, the Time-Dependent Hartree (TDH) approximation \cite{Magann,Guiang,Messina_2}, the Multiconfigurational Time-Dependent Hartree (MCTDH) approximation \cite{Schroder_2,Schroder_3,Wang}, and the time-dependent density matrix renormalization group (tDMRG)~\cite{Doria}.  For example, the MCTDH method was applied to simulate the implementation of a CNOT quantum gate in a six-dimensional model of ammonia \cite{Schroder_2}, and the tDMRG method was applied to simulate control of the superfluid-Mott insulator transition~\cite{Doria}.

In this paper, we take an alternative approach by considering the Magnus expansion, a series expansion that can be used to approximate the time evolution operator.  A common implementation of the Magnus expansion is to split the Hamiltonian governing the quantum system into an unperturbed term and a perturbative term, and then apply the expansion in the interaction picture~\cite{Blanes}.  The Magnus expansion has been applied to a variety of problems in physics and chemistry~\cite{Blanes}.  In particular, the first-order Magnus approximation in the interaction picture has been used in the study of molecular alignment and orientation in short laser fields~\cite{Henriksen}, and in the study of molecular collisions in the sudden regime~\cite{Child}.

We develop a methodology for approximate simulations of quantum dynamics in the context of quantum optimal control of interacting systems.  Specifically, we adopt the first-order Magnus approximation in the interaction picture, treating the interactions between different degrees of freedom as the perturbation, and evaluating the expansion only once at the final time $T$.  This approximation can be suitably integrated with gradient-free algorithms for control field optimization; to this end, we combine it with a Stochastic Hill Climbing algorithm.  The methodology is designed for weakly-interacting systems regardless of control field strength, and it has two main computational benefits: (1) it eliminates the need for expensive gradient computations and the evaluation of the system state at intermediate time steps, and (2) the pairwise interactions of common Hamiltonians allows for an accelerated implementation of the first-order Magnus approximation (as described in Section~\ref{Subsubsection_Exploiting_Separability_to_Reduce_Cost_of_Integration}).  We analyze its computational scaling mathematically in the many-body large $N$ limit.  We numerically illustrate the range of problems that the methodology can handle, as well its accuracy, for optimal control of few-body systems.  Specifically, we consider systems of two and three dipole-dipole coupled molecular rotors subject to a global control field.  We explore a few scenarios: the orientation of all rotors in the same direction, the orientation of rotors in opposing directions, and entanglement between rotors.  An adequate model of the interactions is essential for the latter two cases, as those objectives cannot be reached in the zeroth-order limit.  Previously, optimal control of quantum rotors has also been studied using an exact model for one-rotor~\cite{Salomon,Turinici,Yoshida} and two-rotor~\cite{Mishima,Yu,Zhu_Jing} systems, as well as using the TDH approximation for systems of up to nine rotors~\cite{Magann}.

The remainder of this paper is organized as follows.  Section~\ref{Section_Theoretical_Background} provides relevant theoretical background for the methodology we develop.  Section~\ref{Section_General_Methodology} presents our methodology for approximate quantum optimal control of interacting systems based on the first-order Magnus approximation, including both the mathematical formulation and the numerical implementation procedure.  Section~\ref{Section_Computational_Cost_Analysis_and_Scaling} presents an analysis of the computational cost and scaling of our methodology.  Section~\ref{Section_Illustration_with_Dipole_Dipole_Coupled_Molecular_Rotors} gives a theoretical description of the model rotor systems and corresponding control objectives.  Section~\ref{Section_Optimal_Control_Numerical_Simulations} presents the numerical simulation results.  We close in Section~\ref{Section_Conclusions} with conclusions and a discussion of future work.  Additionally, Appendices \ref{Appendix_Detailed_Numerical_Procedure} to \ref{Appendix_Three_Rotor_Symmetries} respectively present numerical/computational details for the first-order Magnus approximation, a discussion of the second-order Magnus approximation, benchmark simulations at varying separations for the rotor systems, and an analytical proof of physical symmetries in the three-rotor system.

\section{Theoretical Background \label{Section_Theoretical_Background}}

In this section, we review the relevant background underlying the methodology we introduce in Section~\ref{Section_General_Methodology}.

\subsection{Quantum Optimal Control Theory \label{Subsection_Outline_of_Quantum_Optimal_Control_Theory}}

Quantum optimal control theory seeks to identify a time-dependent control field $\varepsilon(t), t \in [0,T]$, that maximizes a given objective functional $J[T,\varepsilon(\cdot)]$ at the final time $T$ \cite{Brif,Peirce,Werschnik}.  In principle, $J[T,\varepsilon(\cdot)]$ can depend on the control target (e.g. orientation of a molecule) as well as terms that favor certain field characteristics \cite{Brif}.  Although the methodology developed in Section~\ref{Section_General_Methodology} can be used for any $J$ that depends explicitly on both the system state at the final time and the control field at all times, for clarity we will restrict ourselves to cases where $J$ represents the expected value of an observable $O$ at the final time $T$, i.e.
\begin{equation}
\max_{\varepsilon(\cdot)} J[T] = \max_{\varepsilon(\cdot)} \bra{\Psi(T)}{O}\ket{\Psi(T)},
\end{equation}
where $\ket{\Psi(t)}$ is the solution to the time-dependent Schr\"odinger equation
\begin{equation}
i \hbar \frac{\partial}{\partial t} \ket{\Psi(t)} = H(t) \ket{\Psi(t)} \label{eq_time_dependent_Schroedinger_equation},
\end{equation}
given the initial state $\ket{\Psi(0)}$.  The Hamiltonian $H(t)$ can be written as the sum
\begin{equation}
H(t)= H_0 + V(t), \label{eq_general_controlled_Hamiltonian_decomposition}
\end{equation}
where $H_0$ is the time-independent field-free Hamiltonian, and $V(t)$ is the time-varying potential arising from the applied control field $\varepsilon(t)$.  For example, within the dipole approximation
\begin{equation}
V(t) = -\pmb{\mu} \cdot \pmb{\varepsilon}(t) \label{eq_electric_dipole_approx_control_potential},
\end{equation}
where $\pmb{\mu}$ is the total system dipole moment.

\subsection{Numerically Exact Simulation of the Schr\"odinger Equation \label{Subsection_Numerically_Exact_Simulation_of_the_Time_Dependent_Schrodinger_Equation}}

The exact solution $\ket{\Psi(t)}$ to Eq.~\eqref{eq_time_dependent_Schroedinger_equation} is given by
\begin{equation}
\ket{\Psi(t)} = U(t) \ket{\Psi(0)}, \label{eq_time_evolution_of_state}
\end{equation}
where $U(t) \equiv U(t,0)$ is the time evolution operator from $0$ to $t$, given by
\begin{equation}
U(t) = \mathcal{T}\bigg(\exp(-\frac{i}{\hbar} \int_0^t H(t') d t')\bigg) \label{eq_formal_general_time_evolution_operator},
\end{equation}
where $\mathcal{T}$ is the time ordering operator.  This can be solved numerically by directly substituting the Hamiltonian $H(t)$ into Eq.~\eqref{eq_formal_general_time_evolution_operator}, and then using small time step propagators for the evolution.  For discretization of $[0,T]$ into $n$ equally-spaced, sufficiently-small time steps $\Delta t \equiv \frac{t}{n}$, this yields
\begin{equation}
\ket{\Psi(n \Delta t)} = \Big(\prod_{k=1}^{n} \exp\Big(- \frac{i}{\hbar} H(k \Delta t) \Delta t\Big) \Big) \ket{\Psi(0)},  \label{eq_short_timestep_propagator}
\end{equation}
taken to be in time order, which converges to the exact solution for $n \rightarrow \infty$.  In this paper we will refer to Eq.~\eqref{eq_short_timestep_propagator} with sufficiently large $n$ (see \cite{Kosloff_0}) as the exact simulation.  The numerical implementation of Eq. \eqref{eq_short_timestep_propagator} requires the exponentiation \cite{footnote1} of the matrix $\big( -\frac{i}{\hbar} H(k \Delta t) \Delta t \big)$ at every time step, which can be computationally prohibitive for systems with interacting degrees of freedom.

\subsection{Magnus Expansion in the Interaction Picture \label{Subsection_Magnus_Expansion_in_the_Interaction_Picture}}

Here we discuss solving Eq.~\eqref{eq_time_dependent_Schroedinger_equation} using the Magnus expansion in the interaction picture (see e.g. \cite{Blanes} for more details).  In the interaction picture, the Hamiltonian $H(t)$ is decomposed into an unperturbed term $H^{(0)}(t)$ and a perturbation term $H^{(1)}(t)$, i.e.
\begin{equation}
H(t) = H^{(0)}(t) + H^{(1)}(t).
\end{equation}
The time evolution operator $U(t)$ (see Eq.~\eqref{eq_formal_general_time_evolution_operator}) is then factored into the product
\begin{equation}
U(t) = U^{(0)}(t) U_I(t), \label{eq_factored_time_evolution_operator}
\end{equation}
where
\begin{equation}
U^{(0)}(t) \equiv \mathcal{T} \bigg(\exp(-\frac{i}{\hbar}\int_0^t H^{(0)}(t')dt')\bigg) \label{eq_zeroth_order_unitary}
\end{equation}
is the time evolution operator corresponding to the evolution generated by $H^{(0)}(t)$ alone, and
\begin{equation}
U_I(t) \equiv \mathcal{T} \bigg( \exp(-\frac{i}{\hbar} \int_0^t H_I^{(1)}(t') dt') \bigg) \label{eq_interaction_time_evolution_operator}
\end{equation}
is the interaction picture time evolution operator, with
\begin{equation}
H^{(1)}_I(t) \equiv U^{(0)}(t)^\dagger H^{(1)}(t) U^{(0)}(t) \label{eq_interaction_picture_perturbation_hamiltonian}.
\end{equation}
The Magnus expansion for $U_I(t)$ is given by
\begin{equation}
U_I(t) = \exp(\Omega_I(t)), \label{eq_interaction_picture_propagator_in_terms_of_omega}
\end{equation}
where
\begin{equation}
\Omega_I(t) \equiv \sum_{k=1}^{\infty} \Omega_{I,k}(t), \label{eq_Omega_I_series_expansion}
\end{equation}
with $\Omega_{I,k}(t)$ being the $k$-th order term in the expansion.  The first and second order terms in Eq.~\eqref{eq_Omega_I_series_expansion} are given respectively by
\begin{equation}
\Omega_{I,1}(t) = -\frac{i}{\hbar} \int_{0}^{t} H_I^{(1)}(t_1)dt_1 \label{eq_Omega_1}
\end{equation}
and
\begin{equation}
\Omega_{I,2}(t) = -\frac{1}{2 \hbar^2}\int_{0}^{t} dt_1 \int_{0}^{t_1} dt_2 [H_I^{(1)}(t_1), H_I^{(1)}(t_2)]. \label{eq_Omega_2}
\end{equation}

Convergence conditions of Eq.~\eqref{eq_interaction_picture_propagator_in_terms_of_omega} are given in \cite{Blanes}, but for practical purposes a $k$-th order approximation is given by truncating the series in Eq. \eqref{eq_Omega_I_series_expansion} after the term $\Omega_{I,k}(t)$.  In particular, the zeroth order Magnus approximation is given by
\begin{equation}
U(t) \approx U^{(0)}(t),
\end{equation}
and the first order Magnus approximation is given by
\begin{equation}
U(t) \approx U^{(0)}(t) \exp(\Omega_{I,1}(t)), \label{eq_Magnus_Expansion_1st_Order}
\end{equation}
which amounts to neglecting the time ordering of the interaction propagator $U_I(t)$ in Eq. \eqref{eq_interaction_time_evolution_operator}.  Note that Eq. \eqref{eq_Magnus_Expansion_1st_Order} still contains partial contributions from all orders of $H^{(1)}(t)$ due to its exponential form.  Importantly, finite order truncations of the Magnus expansion preserve unitarity of $U(t)$, which guarantees conservation of total probability \cite{Blanes}.

The first-order Magnus approximation can be calculated with a single propagator (containing an integral over all time steps), by evaluating Eq. \eqref{eq_Magnus_Expansion_1st_Order} at the final time $t=T$ to directly compute the final state from the initial state without computing intermediate states:
\begin{equation}
\ket{\Psi(T)} \approx U^{(0)}(T) \exp(\Omega_{I,1}(T)) \ket{\Psi(0)}, \label{eq_first_order_magnus_final_time_wavefunction}
\end{equation}
cf. Eq. \eqref{eq_time_evolution_of_state}.  We expect this approximation to be appropriate when the perturbation strength $H^{(1)}(t)$ is sufficiently weak and the propagation time $T$ is sufficiently short.

\section{General Methodology \label{Section_General_Methodology}}

In this section, we present our general methodology for approximate optimal control simulations of interacting quantum systems, with a focus on systems with pairwise interactions.  We first describe the mathematical formulation of the approximate dynamics in Section~\ref{Subsection_Approximate_Simulation_of_Weakly_Coupled_Multi_Dimensional_Quantum_Dynamics}, followed by the associated numerical implementation in Section~\ref{Subsection_Numerical_Implementation_of_Our_General_Methodology}.  We then explain how to combine the quantum dynamics approximation with gradient-free algorithms for control field optimization in Section~\ref{Subsection_Integrating_Gradient_Free_Optimization_with_Magnus}.

\subsection{Dynamics of Interacting Quantum Systems in the First-Order Magnus Approximation\label{Subsection_Approximate_Simulation_of_Weakly_Coupled_Multi_Dimensional_Quantum_Dynamics}}

Here we apply the first-order Magnus approximation in the interaction picture, described in Section~\ref{Subsection_Magnus_Expansion_in_the_Interaction_Picture}, to interacting quantum systems under the influence of a control field.

We consider quantum systems with $N$ degrees of freedom governed by a Hamiltonian of the form
\begin{equation}
\begin{aligned}
H(q_1,q_2,...,q_N,t)& = \sum_{i=1}^N \Big( T(q_i) + V_i(q_i,t) \Big) \\
&+ W(q_1,q_2,...,q_N), \label{eq_fully_general_many_body_Hamiltonian}
\end{aligned}
\end{equation}
where $q_i$ is the coordinate for the $i$-th degree of freedom, $T_i(q_i)$ is the kinetic term for the $i$-th degree of freedom, $V_i(q_i,t)$ is the local potential that acts only on the $i$-th degree of freedom, and $W(q_1,q_2,...,q_N)$ contains the interactions that couple different degrees of freedom (assumed time-independent).  The time-dependence of each $V_i(q_i,t)$ term arises as a result of an external control field $\varepsilon(t)$.

In this paper, we are concerned with the regime where the interactions $W(q_1,q_2,...q_N)$ are weak and can be treated as a perturbation.  As such, we begin by decomposing the Hamiltonian (Eq.~\eqref{eq_fully_general_many_body_Hamiltonian}) into a perturbation term
\begin{equation}
H^{(1)}(q_1,q_2,...,q_N) = W(q_1,q_2,...,q_N) \label{eq_first_order_Hamiltonian}
\end{equation}
and an unperturbed Hamiltonian
\begin{equation}
H^{(0)}(q_1,q_2,...,q_N,t) = \sum_{i=1}^N 
H_i(q_i,t), \label{eq_zeroth_order_Hamiltonian}
\end{equation}
where
\begin{equation}
H_i(q_i,t) \equiv T_i(q_i) + V_i(q_i,t)
\end{equation}
is the coupling-free Hamiltonian for the $i$-th degree of freedom ($i = 1,2,...,N$).  We then apply the first-order Magnus approximation in the interaction picture by substituting into Eq.~\eqref{eq_first_order_magnus_final_time_wavefunction}.

In this particular scenario $H^{(1)}$ is time-independent, and $H^{(0)}(t)$ is a sum of time-dependent terms $H_i(q_i,t)$ that each acts on only a single degree of freedom $i$.  As a result, the zeroth-order time evolution operator $U^{(0)}(q_1,q_2,...,q_N;t)$ (cf. Eq.~\ref{eq_zeroth_order_unitary}) is fully separable into
\begin{equation}
U^{(0)}(q_1,q_2,...;t) = \bigotimes_{i=1}^N U_i(q_i,t), \label{eq_separable_zeroth_order_unitary}
\end{equation}
where $\bigotimes$ denotes a tensor product and
\begin{equation}
U_i(q_i,t) \equiv \mathcal{T}\bigg(\exp(-\frac{i}{\hbar} \int_0^t H_i(q_i,t') d t')\bigg) \label{eq_coupling_free_time_evolution_operator}
\end{equation}
is the coupling-free time evolution operator associated with $H_i(q_i,t)$ ($i = 1,2,...,N$).

Therefore, Eq.~\eqref{eq_first_order_magnus_final_time_wavefunction} can be expressed as
\begin{equation}
\ket{\Psi(T)} \approx \big( \bigotimes_{i=1}^N U_i(q_i,T) \big) \exp(\Omega_{I,1}(T)) \ket{\Psi(0)}, \label{eq_first_order_magnus_approximation_applied_to_interacting_system}
\end{equation}
where
\begin{equation}
\begin{aligned}
\Omega_{I,1}(T) = -\frac{i}{\hbar} \int_{0}^{T} dt' &\big( \bigotimes_{i=1}^N U_i(q_i,t')^\dagger \big) W(q_1, q_2,...,q_N) \\
&\times \big( \bigotimes_{i'=1}^N U_{i'}(q_{i'},t') \big). \label{eq_omega_I_1_for_interacting_systems}
\end{aligned}
\end{equation}

Our choice of how to decompose $H$ into $H^{(0)}$ and $H^{(1)}$ contrasts with \cite{Henriksen} and \cite{Demiralp}, which both choose $V(t)$, as defined in Eq.~\eqref{eq_electric_dipole_approx_control_potential}, as the perturbation.  Our choice is intended to address two important, yet competing, goals: (1) good accuracy beyond the weak-field limit and (2) low computational cost.  The former is justified because for weak interactions, the first-order Magnus approximation is likely a reasonable approximation to the full Magnus expansion.  This is especially true when the total time $T$ is also short, since this reduces the effects of neglecting time-ordering.  Additionally, including the effects of the control field in $H^{(0)}$ allows for strong control fields $\varepsilon(t)$ without compromising accuracy.  The latter is justified because $U^{(0)}$ is fully separable, cf. Eq. \eqref{eq_separable_zeroth_order_unitary}.  This allows for a fast computation of $U^{(0)}(t)$; it also allows for a fast computation of $\Omega_{I,1}(T)$ if $W$ consists only of pairwise terms (cf. Section~\ref{Section_Computational_Cost_Analysis_and_Scaling}).

This mathematical formulation can also be adapted (with greater computational cost) to include higher order terms in the Magnus expansion (Eq.~\eqref{eq_Omega_I_series_expansion}) or a multi-step Magnus scheme  \cite{footnote2}.  In particular, in Appendix~\ref{Appendix_Second_Order_Magnus_Approximation}, we describe the second-order Magnus approximation applied to interacting quantum systems, including the mathematical formulation, a numerical procedure, and a bound on computational cost.

\subsection{Numerical Implementation \label{Subsection_Numerical_Implementation_of_Our_General_Methodology}}

Here we present the procedure for numerically implementing the formulation described in Section~\ref{Subsection_Approximate_Simulation_of_Weakly_Coupled_Multi_Dimensional_Quantum_Dynamics}.  We first outline the procedure for arbitrary multi-degree of freedom Hamiltonians, and then overview a specific implementation with reduced computational cost for Hamiltonians with generic \textit{pairwise} interactions.
 Implementation details as well as a computational cost analysis are presented in Appendix~\ref{Appendix_Detailed_Numerical_Procedure}.  In what follows, the coordinate arguments are suppressed for conciseness.

\subsubsection{General Implementation \label{Subsubsection_Fully_Generic_Numerical_Procedure}}

Here we present a three-part procedure for numerically implementing the first-order Magnus approximation (i.e. Eq.~\eqref{eq_first_order_magnus_approximation_applied_to_interacting_system}), for the multi-degree of freedom Hamiltonian given in Eq.~\eqref{eq_fully_general_many_body_Hamiltonian}.  We discretize the time uniformly into $n$ time steps of size $\Delta t = \frac{T}{n}$ (where $\Delta t$ is sufficiently small compared to the characteristic time scales of the system, see e.g. \cite{Kosloff_0}).

\underline{\textit{Part I:}} $U_i(k \Delta t)$ (see Eq.~\eqref{eq_coupling_free_time_evolution_operator}) is computed, for all $1 \leq i \leq N$ and $1 \leq k \leq n$, recursively starting with $U_i(0) = I$:
\begin{equation}
U_i(k \Delta t) = \exp(-\frac{i}{\hbar} H_i(k \Delta t) \Delta t) U_i((k-1) \Delta t), \label{eq_discretized_coupling_free_time_evolution_op}
\end{equation}
where the short-time propagator $\exp(-\frac{i}{\hbar} H_i(k \Delta t) \Delta t)$ is invoked at each time step (here from the $(k-1)$-th step to the $k$-th step).

\underline{\textit{Part II:}} Using the results of Part I, $\Omega_{I,1}(T)$ is computed, for $T = n\Delta t$, by numerically integrating Eq. \eqref{eq_omega_I_1_for_interacting_systems} from $t = 0$ to $t = T$, i.e.:
\begin{equation}
\Omega_{I,1}(n \Delta t) = -\frac{i}{\hbar} \sum_{k=1}^n \Delta t \big(\bigotimes_{i=1}^N U_i(k \Delta t)^\dagger \big) W \big(\bigotimes_{i'=1}^N U_{i'}(k \Delta t) \big), \label{eq_numerical_integration_multi_body_omega_1}
\end{equation}
where we have invoked the $n$-point rectangular quadrature for the integration with respect to time $t$ over the interval $[0,T]$.  For Hamiltonians whose interactions are all pairwise, Eq.~\eqref{eq_numerical_integration_multi_body_omega_1} can be rearranged to allow for an implementation with reduced computational cost (see Section~\ref{Subsubsection_Exploiting_Separability_to_Reduce_Cost_of_Integration}).

\underline{\textit{Part III:}} Using the results of Part I and II, the final state at time $T = n \Delta t$ is computed as per Eq.~\eqref{eq_first_order_magnus_approximation_applied_to_interacting_system}, by evaluating the approximate equation:
\begin{equation}
\ket{\Psi(n \Delta t)} \approx \big(\bigotimes_{i=1}^N U_i(n \Delta t) \big) \exp(\Omega_{I,1}(n \Delta t)) \ket{\Psi(0)}. \label{eq_approximate_final_state_numerical_eval}
\end{equation}

This implementation of the first-order Magnus approximation computes the final state without explicitly computing the state $\ket{\Psi(k \Delta t)}$ or time evolution operator $U(\Delta t, (k-1) \Delta t)$ at each intermediate time step.  Instead, a numerical integral is carried out over all of the time steps, i.e. Eq.~\eqref{eq_numerical_integration_multi_body_omega_1} in Part II.  As such, the first-order Magnus approximation is particularly computationally convenient when combined with gradient-free optimization methods, for which the objective functional evaluated only at the final time $T$ is typically all that is called for.

\subsubsection{Implementation for Hamiltonians with Pairwise Interactions \label{Subsubsection_Exploiting_Separability_to_Reduce_Cost_of_Integration}}

For many physical quantum systems, the Hamiltonian's  interaction terms are  all pairwise. It is possible to exploit this property, together with the fact that $U^{(0)}(t)$ is always fully separable (see Eq. \eqref{eq_separable_zeroth_order_unitary}), to reduce the cost of computing $\Omega_{I,1}(T)$ in Part II of Section \ref{Subsubsection_Fully_Generic_Numerical_Procedure}.

Here we consider the general case of pairwise interactions of the form
\begin{equation}
W = \sum_{1 \leq i < j \leq N} W_{ij}, \label{eq_sum_of_pairwise_interactions}
\end{equation}
where $W_{ij}$ couples the $i$-th and $j$-th degrees of freedom.  We assume for simplicity that the state of each individual degree of freedom is represented using $D$ dimensions.  In what follows, we will denote the finite-dimensional computational representation of the $i$-th state space as $\mathcal{\tilde{H}}_i$, which is a suitably truncated version of the true Hilbert space $\mathcal{H}_i$ if $\mathcal{H}_i$ is infinite-dimensional.  We will also use the notation $\mathcal{\tilde{H}} \equiv \bigotimes_{i=1}^N \mathcal{\tilde{H}}_i$.

Computing $\Omega_{I,1}(T)$ requires evaluating the numerical integral in Eq. \eqref{eq_numerical_integration_multi_body_omega_1}.  Without exploiting the pairwise property of $W$ given in Eq.~\eqref{eq_sum_of_pairwise_interactions}, the naive approach is to directly multiply together the $D^N \times D^N$ dimensional matrices $\bigotimes_{i=1}^N U_i(k \Delta t)^\dagger$, $W$ and $\bigotimes_{i'=1}^N U_{i'}(k \Delta t)$ at each time step, and then sum the results from all time steps (see Appendix \ref{Appendix_Subsection_Numerical_Procedure_Part_2} for details).

When $W$ assumes the pairwise form given in Eq. \eqref{eq_sum_of_pairwise_interactions}, we can rewrite Eq. \eqref{eq_numerical_integration_multi_body_omega_1} by substituting Eq. \eqref{eq_sum_of_pairwise_interactions} into Eq. \eqref{eq_numerical_integration_multi_body_omega_1} and making some further rearrangements, as follows:
\begin{equation}
\Omega_{I,1}(T) =  \sum_{1 \leq i < j \leq N} \gamma_{ij}, \label{eq_sum_of_gamma_ij}
\end{equation}
where
\begin{equation}
\gamma_{ij} \equiv -\frac{i}{\hbar} \Delta t \sum_{k = 1}^n \big(\bigotimes_{i'=1}^N U_{i'}(k \Delta t)^\dagger \big) W_{ij} \big(\bigotimes_{i''=1}^N U_{i''}(k \Delta t) \big). \label{eq_gamma_ij_before_recasting}
\end{equation}
Since $W_{ij}$ acts non-trivially only on the $i$-th and $j$-th degrees of freedom, Eq.~\eqref{eq_gamma_ij_before_recasting} can be cast as
\begin{equation}
\begin{aligned}
\gamma_{ij} = -\frac{i}{\hbar} \Delta t \sum_{k = 1}^n &\big(U_{i}(k \Delta t)^\dagger \otimes U_j(k \Delta t)^\dagger \big) W_{ij}\\
&\times\big(U_{i}(k \Delta t) \otimes U_j(k \Delta t) \big), \label{eq_gamma_ij}
\end{aligned}
\end{equation}
where the tensor products with identity are suppressed.  Hence, each $\gamma_{ij}$ only contains terms that act non-trivially only on the $i$-th and $j$-th degrees of freedom, and so we can evaluate Eq. \eqref{eq_gamma_ij} for each $(i,j)$ within the $D^2$-dimensional state space $\mathcal{\tilde{H}}_i \otimes \mathcal{\tilde{H}}_j$.  At the end, we add together all of the $\gamma_{ij}$ to get $\Omega_{I,1}(T)$.  

A detailed numerical implementation as well as an analysis comparing the computational costs of the naive method and the method leveraging pairwise interactions are presented in Appendix~\ref{Appendix_Subsection_Numerical_Procedure_Part_2}.  The result is that leveraging pairwise interactions reduces the complexity of Part II of the numerical procedure from exponential to polynomial scaling with respect to $N$.  This improvement arises from the fact that the computing $\Omega_{I,1}(T)$ in this manner eliminates the need to compute products in the full $D^N$-dimensional state space, together with the fact that there are at most $\frac{N(N-1)}{2}$ distinct $\gamma_{ij}$.  Additionally, we note that analogous approaches for exploiting separability can also be developed for an interaction term $W$ that features sums of three-body interactions, sums of four-body interactions, etc.

\subsection{Control Field Optimization within the First-Order Magnus Approximation Framework \label{Subsection_Integrating_Gradient_Free_Optimization_with_Magnus}}

For optimal control simulations with the first-order Magnus approximation, it is necessary to combine our quantum dynamics approximation (see Sections \ref{Subsection_Approximate_Simulation_of_Weakly_Coupled_Multi_Dimensional_Quantum_Dynamics} and \ref{Subsection_Numerical_Implementation_of_Our_General_Methodology}) with an optimization algorithm that does not require knowledge of the system state or time evolution operator at intermediate times, such as a gradient-free algorithm.  For this purpose we adopt a Stochastic Hill Climbing algorithm~\cite{Magann,Rosete_Suarez} as an example algorithm.

\subsubsection{Stochastic Hill Climbing Optimization Algorithm \label{Subsubsection_Stochastic_Hill_Climbing_Optimization_Algorithm}}

In this section we denote $J(T)$ as $J$.  We define $H_{\varepsilon}(t)$ as the Hamiltonian associated with the control field $\varepsilon(t)$, $\varepsilon_{best}(t)$ as the best control field found so far in the optimization, $J_{best}$ as the corresponding objective value, and $J_{thresh}$ as the desired objective value.  The control field $\varepsilon(t)$ is optimized iteratively as follows:
\begin{enumerate}
\item Start with a trial field $\varepsilon_{\textrm{trial}}(t)$.
\item Compute $\ket{\Psi_{\textrm{approx}}(T)}$ from the initial state $\ket{\Psi(0)} = \ket{\Psi_0}$ using the first-order Magnus approximation (Eq.~\eqref{eq_first_order_magnus_approximation_applied_to_interacting_system}), with the Hamiltonian $H_{\varepsilon_{\textrm{trial}}}(t)$.  Then compute the corresponding objective functional $J_{\textrm{trial}}$.  Set $\varepsilon_{\textrm{best}}(t)$ to $\varepsilon_{\textrm{trial}}(t)$ and $J_{\textrm{best}}$ to $J_{\textrm{trial}}$.
\item Suitably choose a sequence of small random changes $\delta \varepsilon(t_k)$ for each time step $k$, where $t_k = k \Delta t$ ($k = 1, 2,... , n$; $n \Delta t = T$).

\item Compute $\ket{\Psi_{\textrm{approx}}(T)}$ from the same initial state $\ket{\Psi_0}$ using the first-order Magnus approximation with the Hamiltonian $H_{\varepsilon_{\textrm{best}} + \delta \varepsilon}(t)$.  Then compute the corresponding objective functional $J_{\varepsilon_{\textrm{best}} + \delta \varepsilon}$.
\item If $J_{\varepsilon_{\textrm{best}} + \delta \varepsilon} > J_{\textrm{best}}$, then update $J_{\textrm{best}}$ to $J_{\varepsilon_{\textrm{best}} + \delta \varepsilon}$ and $\varepsilon_{\textrm{best}}(t)$ to $\varepsilon_{\textrm{best}}(t) + \delta \varepsilon(t)$.  Otherwise, do not update $J_{\textrm{best}}$ or $\varepsilon_{\textrm{best}}(t)$.
\item If $J_{\textrm{best}} \geq J_{\textrm{thresh}}$, then terminate the optimization and set the optimized field as $\varepsilon_{\textrm{opt}}(t) = \varepsilon_{\textrm{best}}(t)$.  Otherwise, return to step 3.
\end{enumerate}

\section{Computational Cost and Scaling\label{Section_Computational_Cost_Analysis_and_Scaling}}

The total computational cost of solving a quantum optimal control problem is mainly determined by two factors: (1) the number of iterations to converge to the optimal field and (2) the cost per iteration.  The latter takes into account the cost of simulating the whole dynamics for $t \in [0,T]$, plus some overhead cost (for example, generating the random changes $\delta \varepsilon(t)$ in the control field $\epsilon(t)$).  Here we restrict our analysis to the cost per iteration only (ignoring the overhead cost), for systems with pairwise interactions (i.e. Eq.~\eqref{eq_sum_of_pairwise_interactions}).

We analyze the computational cost and scaling by counting the number of complex number multiplications, as is conventional for estimating the cost of matrix computations~\cite{Moler,footnote3}.  In our notation, we use \textit{big} $O$ to denote an upper bound on computational complexity (up to a constant), and when we do not use the \textit{big} $O$ we denote the exact computational cost using the currently-known best methods.   For simplicity, we assume that the state space of each individual degree of freedom is represented by the same number of dimensions $D$.  Additionally, we make use of the fact that the computational cost of computing $e^A \mathbf{v}$ given a $d \times d$ matrix $A$ and a $d$-dimensional vector $\mathbf{v}$ is $C_1 d^{\alpha}$, where $C_1$ and $2 < \alpha \leq 3$ are constants (note this computation can be performed without explicit computation of $e^A$)~\cite{Al_Mohy,Moler,Higham}.  Henceforward, we will refer to this operation as a "$d$-dimensional computation of the form $e^A \mathbf{v}$".

For the first-order Magnus approximation, the computational cost of each part of our numerical implementation (outlined in Section~\ref{Subsection_Numerical_Implementation_of_Our_General_Methodology}) is analyzed in Appendix~\ref{Appendix_Detailed_Numerical_Procedure}.  In brief, the cost of Part I is $O(n N D^3)$, the cost of Part II is $O(n N^2 D^6)$, and the cost of Part III is $C_1 D^{\alpha N} + 2 D^{2N} + O(D^{N+1})$.  Therefore, the total computational cost of one whole simulation of the dynamics can be calculated as $C_1 D^{\alpha N} + 2 D^{2N} + O(D^{N+1}) + O(n N^2 D^6)$. 

The actual costs will be highly problem-dependent, so we will now compare the relative costs of the first-order Magnus approximation with the exact simulation, in the large $N$ limit.  In this limit, the dominant cost of the first-order Magnus approximation is $C_1 D^{\alpha N}$, which arises exclusively from a single $D^N$-dimensional computation of the form $e^A \mathbf{v}$ (i.e. $\exp(\Omega_{I,1}(T)) \ket{\Psi(0)}$, cf. Eq.~\eqref{eq_approximate_final_state_numerical_eval}).  The dominant cost of the exact simulation is $C_1 n D^{\alpha N}$, which arises exclusively from $n$ total $D^N$-dimensional computations of the form $e^A \mathbf{v}$ (i.e. $\exp(-\frac{i}{\hbar}\Delta t H(k \Delta t)) \ket{\Psi((k-1) \Delta t)}$ for all $k=1,2,...n$, cf. Eq.~\eqref{eq_short_timestep_propagator}).  Therefore, at large $N$, the computational cost of the first-order Magnus approximate simulation is roughly $\frac{1}{n}$ times the cost of the exact simulation~\cite{footnote4}. This is a significant improvement since the number of time steps $n$ is generally a very large number~\cite{Kosloff_0} (for reference, $n$ is on the order of $10^3$ to $10^4$ for our numerical simulations in Section~\ref{Section_Optimal_Control_Numerical_Simulations}).  Moreover, it is evident that this $\frac{1}{n}$ computational improvement of the first-order Magnus approximation arises from not computing the system state at the intermediate time steps, which is enabled by the fact for a gradient-free optimization algorithm knowledge of the intermediate states is unnecessary.

For truly large $N$, eventually both the first-order Magnus approximation and the exact simulation are computationally intractable, and so we should concern ourselves with how realistic the $\frac{1}{n}$ improvement estimate is for practical problems.  An important practical issue is pushing the upper limit on the number of degrees of freedom $N$ for which the simulation is not computationally prohibitive.  In the regime of $N$ near this upper limit, since intractability arises from exponential scaling with respect to $N$, it is reasonable to assume that the terms $O(D^{N+1}) + O(n N^2 D^6)$ are dominated by the terms $C_1 D^{\alpha N} + 2 D^{2N}$.  However, it is less clear whether $2 D^{2N}$ is truly negligible.  Nevertheless, even accounting for these two terms, the order of magnitude of the $\frac{1}{n}$ estimate is not likely to be significantly changed (similarly, adding in the cost of actually computing the objective functional $J$ given the final wavefunction is unlikely to affect the order of magnitude of the estimate).  Additionally, there are other costs in the exact simulation other than the $n$ $D^N$-dimensional computations of the form $e^A \mathbf{v}$ that may be non-negligible for practical problems, and accounting for those could potentially give a more favorable scaling estimate.

Additionally, in Appendix \ref{Appendix_Second_Order_Magnus_Approximation}, we show that, still assuming $W$ has the pairwise form given in Eq.~\eqref{eq_sum_of_pairwise_interactions}, the total computational cost of one whole simulation of the second-order Magnus approximation will be $C_1 D^{\alpha N} + 2 D^{2N} + O(D^{N+1}) + O(n^2 N^3 D^9)$.  At sufficiently large $N$, the term $C_1 D^{\alpha N}$ will clearly dominate the term $O(n^2 N^3 D^9)$.  As such, at sufficiently large $N$, the dominant computational costs of the second-order and first-order Magnus approximations will be the same, and the cost of simulating the approximate dynamics using the second-order Magnus approximation will also be roughly $\frac{1}{n}$ times the cost of simulating the exact dynamics.  However, our upper bound on complexity for computing $\Omega_{I,2}(T)$ ($O(n^2 N^3 D^9)$) is significantly greater than our bound for computing $\Omega_{I,1}(T)$ ($O(n N^2 D^6)$), and so further analysis is needed to determine whether the second-order Magnus approximation will be useful for practical-sized problems.

The above analysis show that efforts to further reduce the computational cost of our numerical implementation for many-body systems should focus on the evaluation of Eq. \eqref{eq_approximate_final_state_numerical_eval} in Part III (since Parts I and II scale as a polynomial with respect to $N$).  For example, it may be possible to reduce computational cost by taking advantage of matrix/vector sparsity or system-specific physical symmetries~\cite{Gilbert,Lanczos,Sidjea}.  Additionally, one interesting feature of the Stochastic Hill Climbing approach is that we expect both $U_i(T)$ (for each $i$) and $\Omega_{I,1}(T)$ in Eq. \eqref{eq_approximate_final_state_numerical_eval} to only be slightly changed between successive iterations of the optimization algorithm; we leave open the question of whether it is possible to leverage on this feature to reduce computational cost of successive iterations.

\section{Theoretical Description of Rotor System and Corresponding Control Objectives \label{Section_Illustration_with_Dipole_Dipole_Coupled_Molecular_Rotors}}

\subsection{System Description \label{Subsection_System_Description}}

\subsubsection{General $N$-Rotor System}

\begin{figure}
\centering
\begin{subfigure}{0.275\textwidth}
\centering
\includegraphics[width=1.0\linewidth]{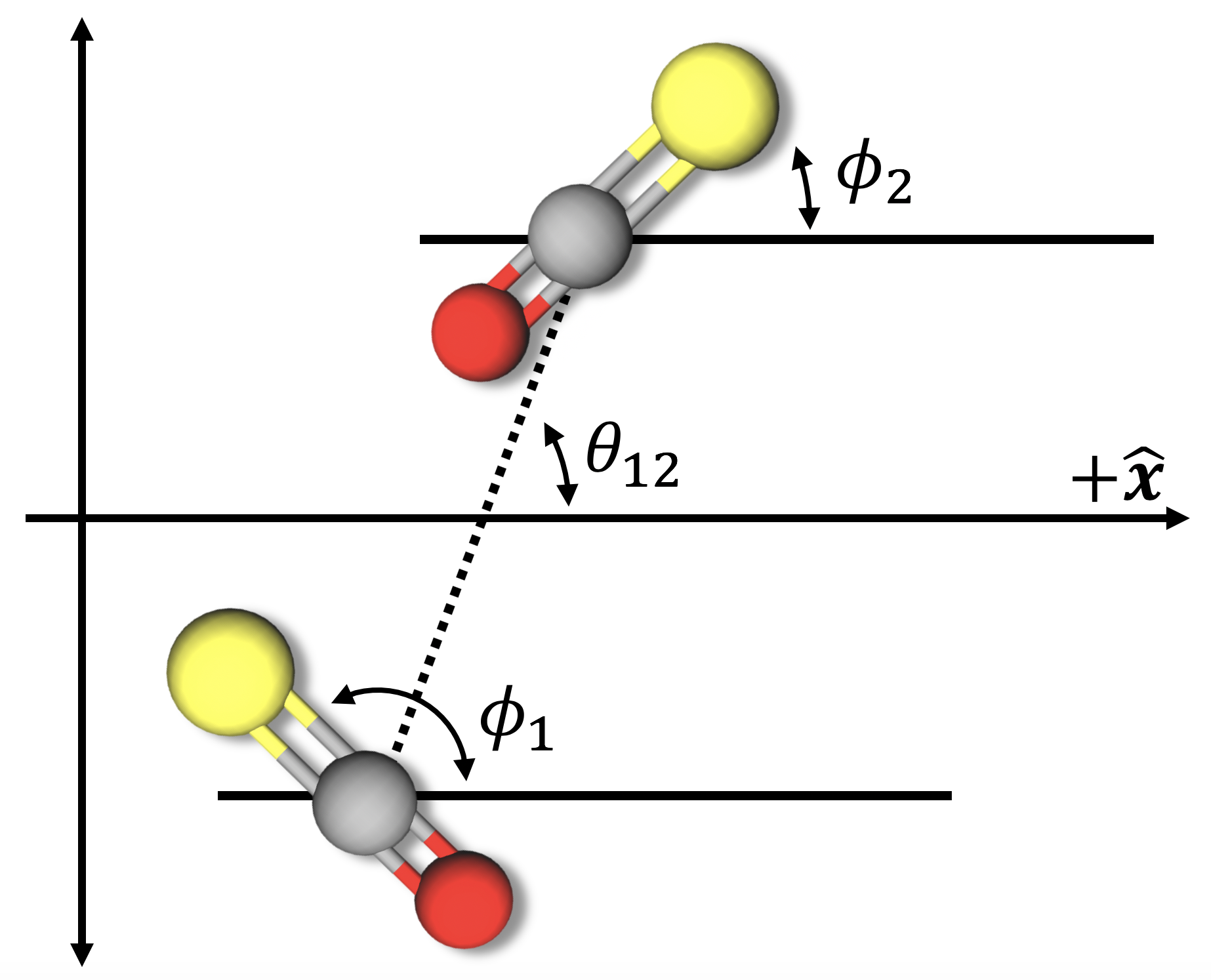}
\caption{Two-rotor schematic.} \label{fig_2_rotor_schematic}
\end{subfigure}
\begin{subfigure}{0.4\textwidth}
\centering
\includegraphics[width=1.0\linewidth]
{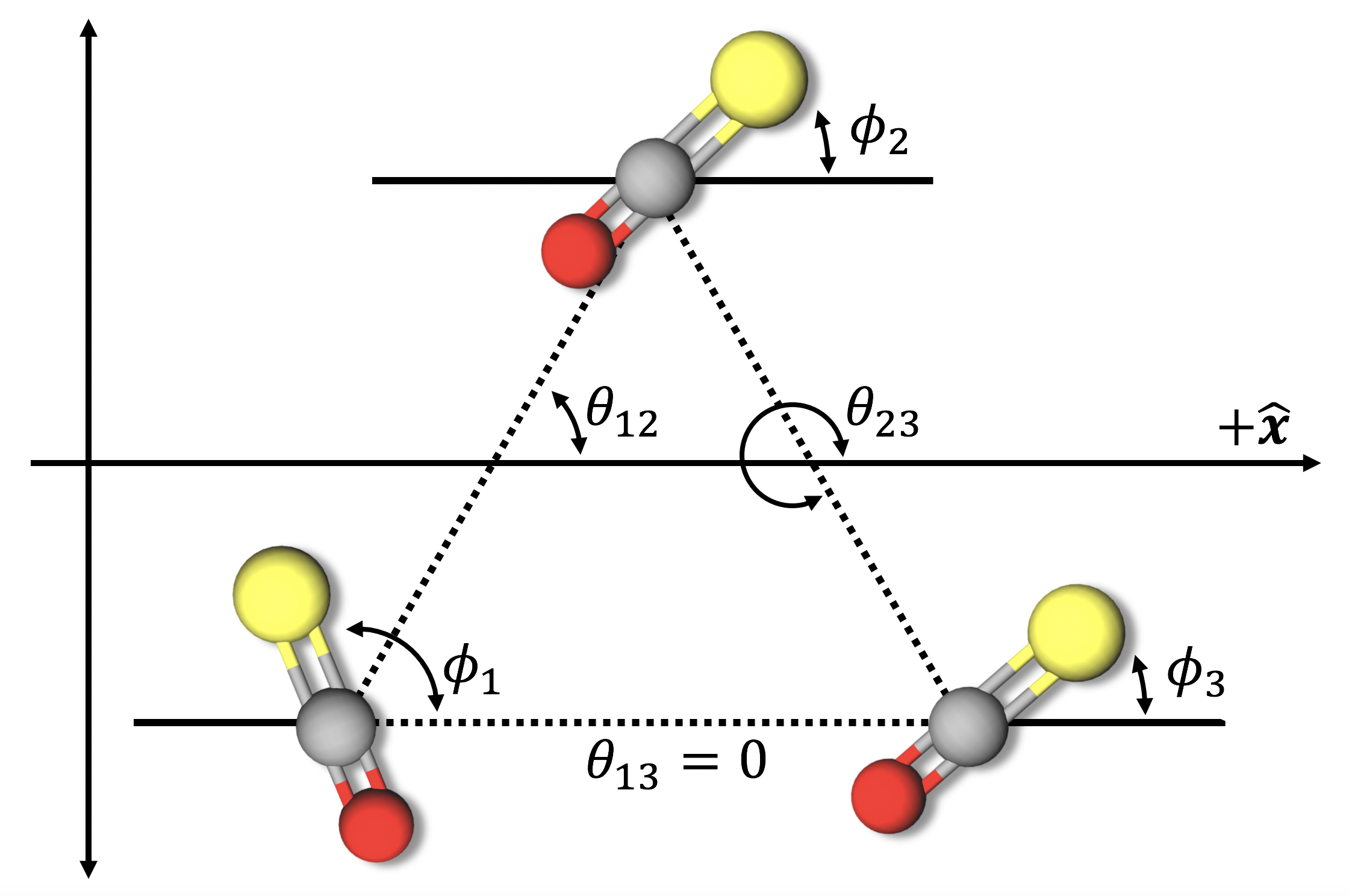}
\caption{Three-rotor schematic.}
\label{fig_3_rotor_schematic}
\end{subfigure}
\caption{(a) Schematic of two coupled OCS molecules.  The rotors are under the influence of a time-dependent control electric field $\varepsilon$ polarized in the $\hat{\textbf{x}}$-direction.  $\phi_1$ and $\phi_2$ are the angles of rotor 1 and rotor 2 with respect to the polarization direction of the electric field.  $\theta_{12}$ is the angle of the vector between rotors 1 and 2 and the polarization direction of the electric field.  (b) Schematic of three coupled OCS molecules.  The rotors are under the influence of a time-dependent control electric field $\varepsilon$ polarized in the $\hat{\textbf{x}}$-direction.  $\phi_j$ for $j=1,2,3$ is the angle of the $j-th$ rotor with respect to the polarization direction of the electric field.  $\theta_{ij}$ for $1 \leq i < j \leq 3$ is the angle between the vector between rotors $i$ and $j$ and the polarization direction of the electric field.  Both figures are reproduced (with modifications) with permission from~\cite{Magann}.} \label{fig_both_rotor_schematics}
\end{figure}

We consider systems of multiple linear OCS (carbonyl sulfide) molecules under the influence of a linearly polarized electric field.  The molecules are modeled as planar rigid rotors that interact via dipole-dipole coupling.  The Hamiltonian for a system of $N$ rotors is given by:
\begin{equation}
\begin{aligned}
H&(\phi_1,...,\phi_N,t) = \sum_{i=1}^N (B L_i^2 - \mu \varepsilon(t) \cos\phi_i) \\
&+ \sum_{1 \leq i < j \leq N} \frac{\mu^2}{4 \pi \epsilon_{0} R_{ij}^3} \Big( \cos(\phi_i - \phi_j)\\
&- 3 \cos(\phi_i - \theta_{ij}) \cos(\phi_j - \theta_{ij}) \Big), \label{eq_general_N_rotor_Hamiltonian}
\end{aligned}
\end{equation}
where $B \equiv \frac{\hbar^2}{2I} = 4.033\times10^{-24} J$ is the rotational constant~\cite{Maki}, $L_i^2 = -\frac{\partial^2}{\partial \phi_i}$ is the angular momentum operator for the $i$-th rotor, $\mu \equiv |\pmb{\mu}_i| = 2.36496 \times 10^{-30}$ C m~\cite{Shulman} is the magnitude of the individual dipole moment of each rotor, $\varepsilon(t)$ is the electric field, $\phi_i$ is the angle between the $i$-th rotor's dipole moment $\pmb{\mu}_i$ and the electric field polarization (in our case oriented in the $\hat{\textbf{x}}$-direction), $\mathbf{R}_{ij}$ is the fixed vector between rotor $i$ and rotor $j$, and $\theta_{ij}$ is the fixed angle between $\mathbf{R}_{ij}$ and the electric field polarization~\cite{Magann,Yu}.  Labeled schematics for $N=2$ and $N=3$ are shown in Figs. \ref{fig_2_rotor_schematic} and \ref{fig_3_rotor_schematic} respectively.

The interactions in the Hamiltonian in Eq.~\eqref{eq_general_N_rotor_Hamiltonian} are all pairwise, and so we can use the numerical implementation described in Section~\ref{Subsubsection_Exploiting_Separability_to_Reduce_Cost_of_Integration}.  In our framework for the first-order Magnus approximation, this corresponds to:

\begin{equation}
H_i(\phi_i,t)= B L_i^2 - \mu \varepsilon(t) \cos\phi_i
\end{equation}
for the coupling-free Hamiltonian and
\begin{equation}
W_{ij} = \frac{\mu^2}{4 \pi \epsilon_{0} R_{ij}^3} \Big( \cos(\phi_i - \phi_j) - 3 \cos(\phi_i - \theta_{ij}) \cos(\phi_j - \theta_{ij}) \Big) \label{eq_W_ij_for_N_rotors}
\end{equation}
for the pairwise interaction between rotors $i$ and $j$.

Following \cite{Magann} and \cite{Yu}, we use the basis set given by \{$\ket{m_1} \otimes \ket{m_2} \cdots \otimes \ket{m_N}$\}, where \{$\ket{m_i}$\} are the orthonormal eigenstates of the angular momentum operator $L_i^2$, satisfying:
\begin{equation}
L_i^2 \ket{m_i} = m_i^2 \ket{m_i},
\end{equation}
where $m_i = -\infty,...,-1,0,1,...,\infty$.  Each $\ket{m_i}$ is given explicitly by
\begin{equation}
\ket{m_i} = \int_0^{2\pi} \ket{\phi_i}\bra{\phi_i}\ket{m_i} d\phi_i,
\end{equation}
where
\begin{equation}
\bra{\phi_i}\ket{m_i} = \sqrt[]{\frac{1}{2 \pi}} e^{i m_i \phi_i}.
\end{equation}

In this basis set, the operators $\cos\phi_i$ and $\sin\phi_i$, respectively, can be represented using the following matrix element relations:
\begin{equation}
\bra{m_i}\cos\phi_i\ket{m_i'} = \frac{1}{2}(\delta_{m_i,m_i'+1} + \delta_{m_i,m'_i-1})
\end{equation}
and
\begin{equation}
\bra{m_i}\sin\phi_i\ket{m_i'} = -\frac{i}{2}(\delta_{m_i,m_i'+1} - \delta_{m_i,m'_i-1}).
\end{equation}

For numerical simulations we truncate to a sufficiently large finite basis of $2M+1$ basis elements for each rotor, such that $m_i = -M,...,-2,-1,0,1,2,...,M$ (so $D = 2M + 1$).

\subsubsection{Two-Rotor Geometry and Initial State \label{Subsubsection_Model_Hamiltonian_for_Two_Coupled_Molecular_Rotors}}

For our two-rotor system simulations (in Section \ref{Subsection_Two_Rotor_System_Optimal_Control_Simulations} and Appendix \ref{Appendix_Accuracy_vs_Separation_of_First_Order_Magnus_Approximation_for_Two_and_Three_Rotor_Simulations}), we specialize to the case of the angle $\theta_{12} = \frac{\pi}{2}$ and the initial state $\ket{\Psi(0)} = \ket{0} \otimes \ket{0}$,
which was also considered in \cite{Yu} (see Fig. \ref{fig_2_rotor_schematic}).  From \cite{Yu}, the symmetries of this choice of $\theta$ makes the two rotors identical even in the presence of the interaction $W_{12}$.  In Appendix \ref{Appendix_Accuracy_vs_Separation_of_First_Order_Magnus_Approximation_for_Two_and_Three_Rotor_Simulations}, we perform benchmark simulations to analyze the accuracy versus separation of the first-order Magnus approximation for this system, and in Section \ref{Subsection_Two_Rotor_System_Optimal_Control_Simulations}, we perform optimal control simulations for this geometry and initial state.  These benchmark simulations indicate that the first-order Magnus approximation is better than the zeroth-order model at all separations, and that the first-order Magnus approximation accuracy improves with increasing separation.

\subsubsection{Three-Rotor Geometry and Initial State \label{Subsubsection_Model_Hamiltonian_for_Three_Coupled_Molecular_Rotors}}

For our three-rotor system simulations (in Section \ref{Subsection_Three_Rotor_Optimal_Control_Simulations} and Appendix \ref{Appendix_Accuracy_vs_Separation_of_First_Order_Magnus_Approximation_for_Two_and_Three_Rotor_Simulations}), we specialize to the case of an equilateral triangle geometry (i.e. $\theta_{12} = \frac{\pi}{3}$, $\theta_{13} = 0$, $\theta_{23} = \frac{5 \pi}{3}$, and $R_{12} = R_{13} = R_{23} \equiv R$) and an initial state given by $\ket{\Psi(0)} = \ket{0} \otimes \ket{0} \otimes \ket{0}$, which was also considered in \cite{Magann} (see Fig. \ref{fig_3_rotor_schematic}).  In Appendix \ref{Appendix_Three_Rotor_Symmetries}, we prove analytically for this three-rotor geometry and initial state, the following relations hold at all times $t$:
\begin{equation}
\langle \cos\phi_1 \rangle(t) = \langle \cos\phi_3 \rangle(t) \label{eq_cosphi1_cosphi_3_relationship},
\end{equation}
\begin{equation}
\langle \sin\phi_1 \rangle(t) = - \langle \sin\phi_3 \rangle(t) \label{eq_sinphi1_sinphi3_relationship},
\end{equation}
and
\begin{equation}
\langle \sin\phi_2 \rangle(t) = 0 \label{eq_sinphi2_relationship},
\end{equation}
for any (not necessarily optimal) field $\varepsilon(t)$, where the time-dependent expectation value of an observable $O$ is defined as $\langle O \rangle(t) \equiv \bra{\Psi(t)}O\ket{\Psi(t)}$.  Note that in general, $\langle \cos\phi_2 \rangle \neq 0$ is allowed.  We remark that the relationships in Eqs. \eqref{eq_cosphi1_cosphi_3_relationship} to \eqref{eq_sinphi2_relationship} indicate that rotors 1 and 3 are non-identical, but $\langle \cos\phi_1 \rangle$ and $\langle \cos\phi_3 \rangle$ behave identically, which will be important for our optimal control simulations.  Benchmark simulations analyzing the accuracy versus separation of the first-order Magnus approximation for this geometry and initial state are presented in Appendix \ref{Appendix_Three_Rotor_Symmetries}, and optimal control simulations for this system are presented in Section \ref{Subsection_Three_Rotor_Optimal_Control_Simulations}.  As in Section~\ref{Subsubsection_Model_Hamiltonian_for_Two_Coupled_Molecular_Rotors}, these benchmark simulations indicate that the first-order Magnus approximation is more accurate at greater separations, and is better than the zeroth-order model at all separations.

\subsection{Control Objectives \label{Subsection_Optimal_Control_Objectives}}

\subsubsection{Two-Rotor Objectives \label{Subsubsection_Two_Rotor_Objectives}}

We define two control objectives for the two-rotor system (see Fig.~\ref{fig_2_rotor_schematic}): (i) identical orientations and (ii) entanglement.

(i) For identical orientations of two rotors, we define the objective functional as
\begin{equation}
J^{[2]}_{id}(T) \equiv \bra{\Psi(T)}\big(\cos\phi_1 + \cos\phi_2 \big)\ket{\Psi(T)} \label{eq_objective_functional_two_rot_id_orientation},
\end{equation}
which corresponds to the expectation value of orientation in the polarization direction of the control field (+$\hat{\textbf{x}}$-direction) for both rotors.

(ii) For entanglement control of two rotors, we define the objective functional as~\cite{Yu}:
\begin{equation}
J^{[2]}_{ent}(T) \equiv |\bra{\Psi(T)}\ket{\Psi^{MES}}|^2, \label{eq_entanglement_objective_functional}
\end{equation}
corresponding to the expected value of the projection operator $\ket{\Psi^{MES}}\bra{\Psi^{MES}}$, where
\begin{equation}
\ket{\Psi^{MES}} \equiv \sqrt[]{\frac{1}{2M + 1}} \sum_{m=-M}^{M} \ket{m} \otimes \ket{m} \label{eq_maximally_entangled_state}
\end{equation}
is a maximally entangled state (for a basis set truncated at $M$). The state  $\ket{\Psi^{MES}}$ maximizes the von Neumann entropy $S_{vN}$~\cite{Amico,Nielsen}, such that
\begin{equation}
S_{vN}(\ket{\Psi^{MES}}) = \ln(2M + 1). \label{eq_maximum_possible_entropy}
\end{equation}

\subsubsection{Three-Rotor Objectives \label{Subsubsection_Three_Rotor_Objectives}}

We define two control objectives for the three-rotor system (see Fig~\ref{fig_3_rotor_schematic}): (i) identical orientations and (ii) opposing orientations.

(i) For identical orientations of three rotors, we define the objective functional as
\begin{equation}
J^{[3]}_{id}(T) \equiv \bra{\Psi(T)}(\cos\phi_1 + \cos\phi_2 + \cos\phi_3)\ket{\Psi(T)} \label{eq_objective_functional_three_rotor_idential_orientation},
\end{equation}
which corresponds to the expectation value of orientation in the direction of the polarization of the control field ($+\hat{\textbf{x}}$-direction) for all three rotors.

(ii) For opposing orientations of the three-rotor system, we define the objective functional as
\begin{equation}
J^{[3]}_{opp}(T) = \bra{\Psi(T)}(-\cos\phi_1 + \cos\phi_2)\ket{\Psi(T)} , \label{eq_objective_functional_three_rotor_opposing_orientation}
\end{equation}
which is equivalent to $J^{[3]}_{opp}(T) = \bra{\Psi(T)}(-\cos\phi_3 + \cos\phi_2)\ket{\Psi(T)}$ (cf. Eq. \eqref{eq_cosphi1_cosphi_3_relationship}).  Maximization of this objective corresponds to orienting rotor 2 in the $+ \hat{\textbf{x}}$-direction, and orienting rotors 1 and 3 in the $- \hat{\textbf{x}}$-direction.

\section{Numerical Results for Optimal Control of Rotors \label{Section_Optimal_Control_Numerical_Simulations}}

In this section, we numerically investigate the utility of the first-order Magnus approximation, in conjunction with the Stochastic Hill Climbing optimization algorithm, by performing optimal control calculations for the coupled molecular rotor systems described in Section~\ref{Subsection_System_Description}. To evaluate the quality of our approximation in each case, we apply the optimal field (obtained using the first-order Magnus approximation) to the exact dynamics.

For all simulations, we choose our initial trial field $\varepsilon_{trial}(t)$ to have a Gaussian envelope and contain a combination of $F$ characteristic frequencies of the system ($F \leq M$), i.e.,
\begin{equation}
\varepsilon_{trial}(t) = a_0 \exp(-\frac{(t-\frac{T}{2})^2}{(\frac{T}{2 \sqrt[]{7}})^2}) \sum_{m=0}^{F-1} b_m \cos(\omega_m t), \label{eq_trial_field_general_form}
\end{equation}
where $\omega_m = \frac{B (2m + 1)}{\hbar}$ is the field-free, interaction-free, individual rotor transition frequency from $\ket{m}$ to $\ket{m+1}$, and $a_0$ and the $b_m$ are some constants we select for the respective optimal control calculations.  To reduce computational expense in our optimizations, we first optimized the control field in simpler systems (such as lower $M$, lower $n$, or fewer rotors), and then used the resulting optimal field on the full system.  In what follows, we present only the results of the final optimizations in the full systems.

\subsection{Two-Rotor Optimal Control Simulations \label{Subsection_Two_Rotor_System_Optimal_Control_Simulations}}

\subsubsection{Identical Orientations \label{Subsubsection_Two_Rotor_System_Symmetric_Orientation_Objective}}

\begin{figure}
\centering
\includegraphics[width=0.48\textwidth]{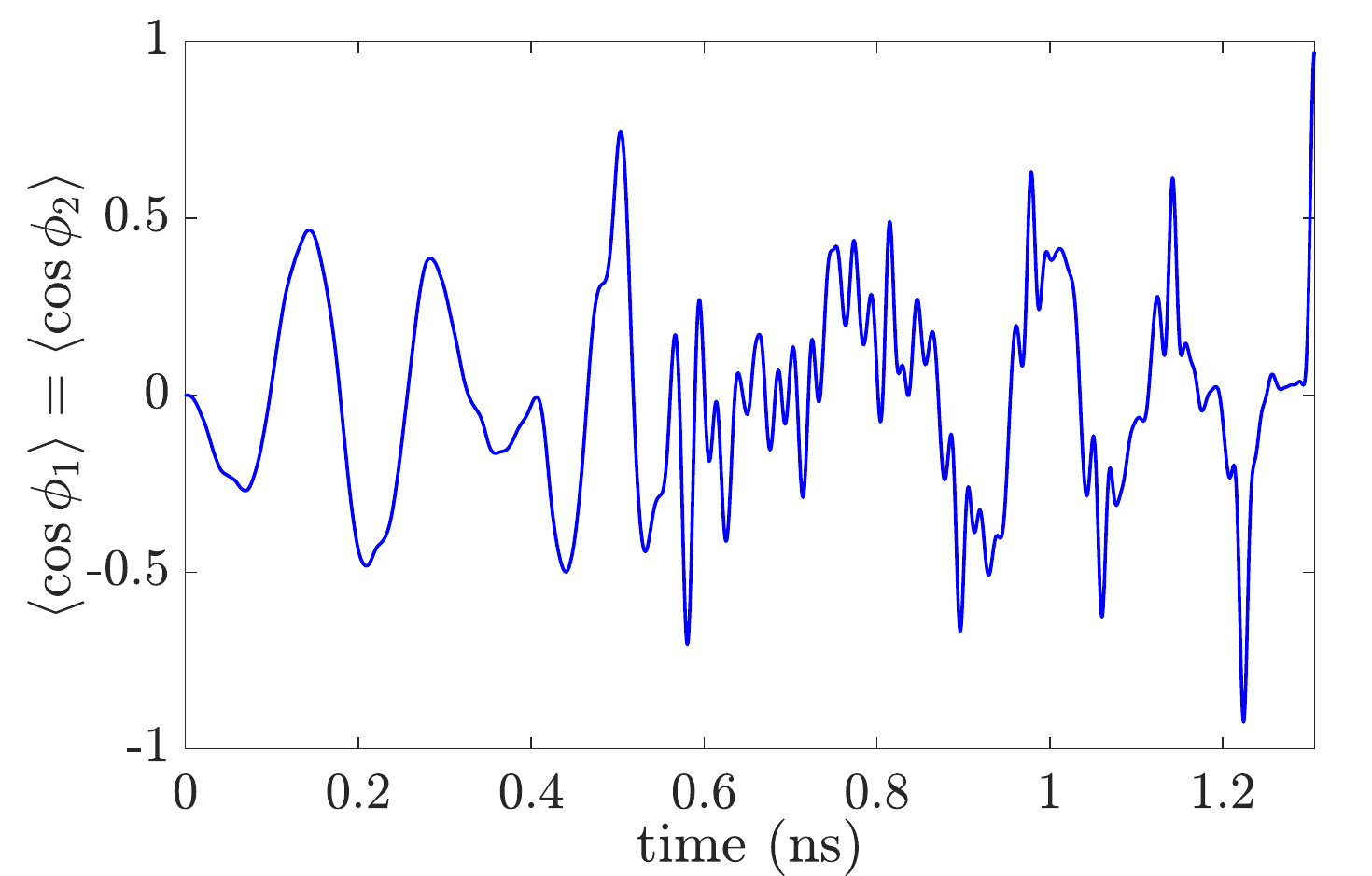}
\caption{Plot of $\langle \cos\phi_1\rangle(t) = \langle\cos\phi_2 \rangle(t)$ vs. time for exact evolution of two-rotor system driven by $\varepsilon(t)$, which is the optimal field obtained from optimization of $J^{[2]}_{id}(T)$ using the first-order Magnus approximation.  We observe a somewhat oscillatory behavior at the start of the time period, and also that the objective rapidly acquires its final value at the end of the time period.}
\label{fig_2_rotor_sym_orientation_vs_time}
\end{figure}

Here we perform optimal control calculations to maximize the objective functional $J^{[2]}_{id}(T)$ defined in Eq.~\eqref{eq_objective_functional_two_rot_id_orientation} for identical orientation of two OCS rotors (see Fig.~\ref{fig_2_rotor_schematic}).  We choose $R = 5$ nm and $T \approx 1.306$ ns, to be approximately consistent with the choice made in \cite{Yu} for exact optimization of the same objective.  Additionally, we chose $M = 8$ for the truncated basis set, $n = 1998$ time steps, and the trial field given by Eq. \eqref{eq_trial_field_general_form}, with $F = 4$, $b_0 = 0.2$, $b_1 = 0.3$, $b_2 = 0.3$, $b_3 = 0.2$, and $a_0 = 8.5625 \times 10^6$ V/m.

The optimal field $\varepsilon(t)$ for $M= 8$ (not shown) yielded $J^{[2]}_{id}(T)$ = 1.96008 within the first-order Magnus approximation.  Using this optimal field on a numerically exact model, we calculated an exact objective of $J^{[2]}_{id}$ = 1.94027.  Hence, we are able to achieve a high value of $J^{[2]}_{id}$, and importantly, we see very little drop-off in objective when using the optimal field found from the first-order Magnus approximate dynamics on the exact dynamics.  For reference, applying this same field to the exact dynamics in the $M=9$ truncated basis yields an exact objective of $J^{[2]}_{id} = 1.94032$, indicating that the $M=8$ model is appropriate for this resultant optimal field.

In Fig.~\ref{fig_2_rotor_sym_orientation_vs_time} the orientation $\langle \cos\phi_1\rangle = \langle \cos\phi_2 \rangle(t)$ is plotted versus time for exact dynamics in the presence of $\varepsilon(t)$, the optimal field we found using the first-order Magnus approximation.  It was found that the orientation starts out with a somewhat oscillatory behavior, and acquires its final optimal value right at the end of the time range.  This is qualitatively similar to the behavior of the optimal field found using the exact solution for the identical orientation of the two-rotor system presented in \cite{Yu}.

\subsubsection{Entanglement \label{Subsubsection_Two_Rotor_System_Entanglement_Objective}}

Here we perform optimal control calculations using the first-order Magnus approximation to maximize the objective functional $J^{[2]}_{ent}(T)$ defined in Eq.~\eqref{eq_entanglement_objective_functional} for entanglement control of two OCS rotors.  We choose $T \approx 3.921$ ns and $R = 7$ nm. Additionally, we chose $M = 4$ for the truncated basis set, $n = 11996$ time steps, and the trial field given by Eq. \eqref{eq_trial_field_general_form}, with $F = 4$, $b_0 = 0.2$, $b_1 = 0.3$, $b_2 = 0.3$, $b_3 = 0.2$, and $a_0 = 4 \times 10^6$ V/m.

\begin{figure}[h!]
\centering
\begin{subfigure}[b]{0.48\textwidth}
\includegraphics[width=\textwidth]{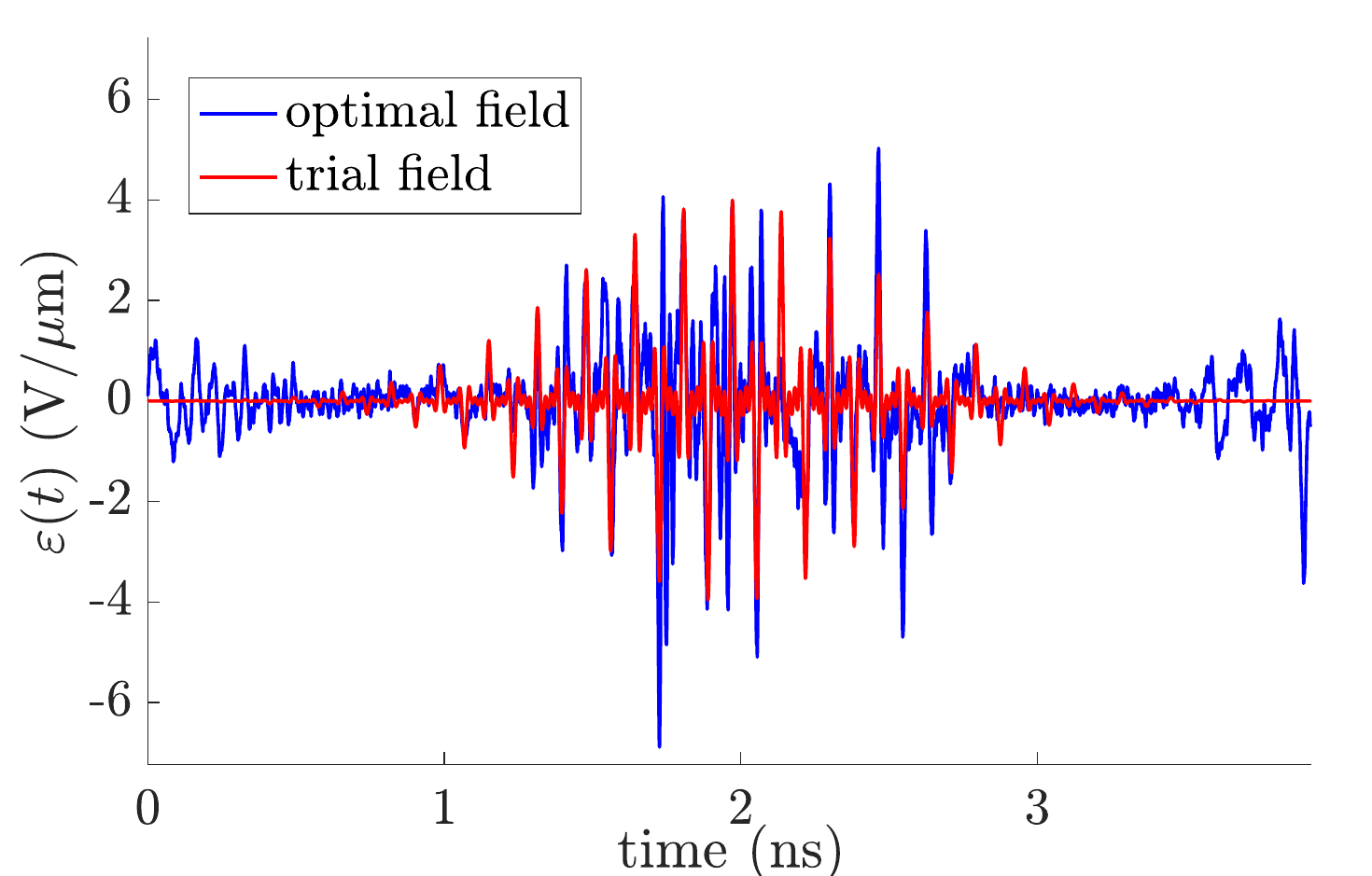}
\caption{Electric Field vs. Time} \label{fig_2_rotor_ent_temporal_field}
\end{subfigure}
\begin{subfigure}[b]{0.48\textwidth}
\includegraphics[width=\textwidth]
{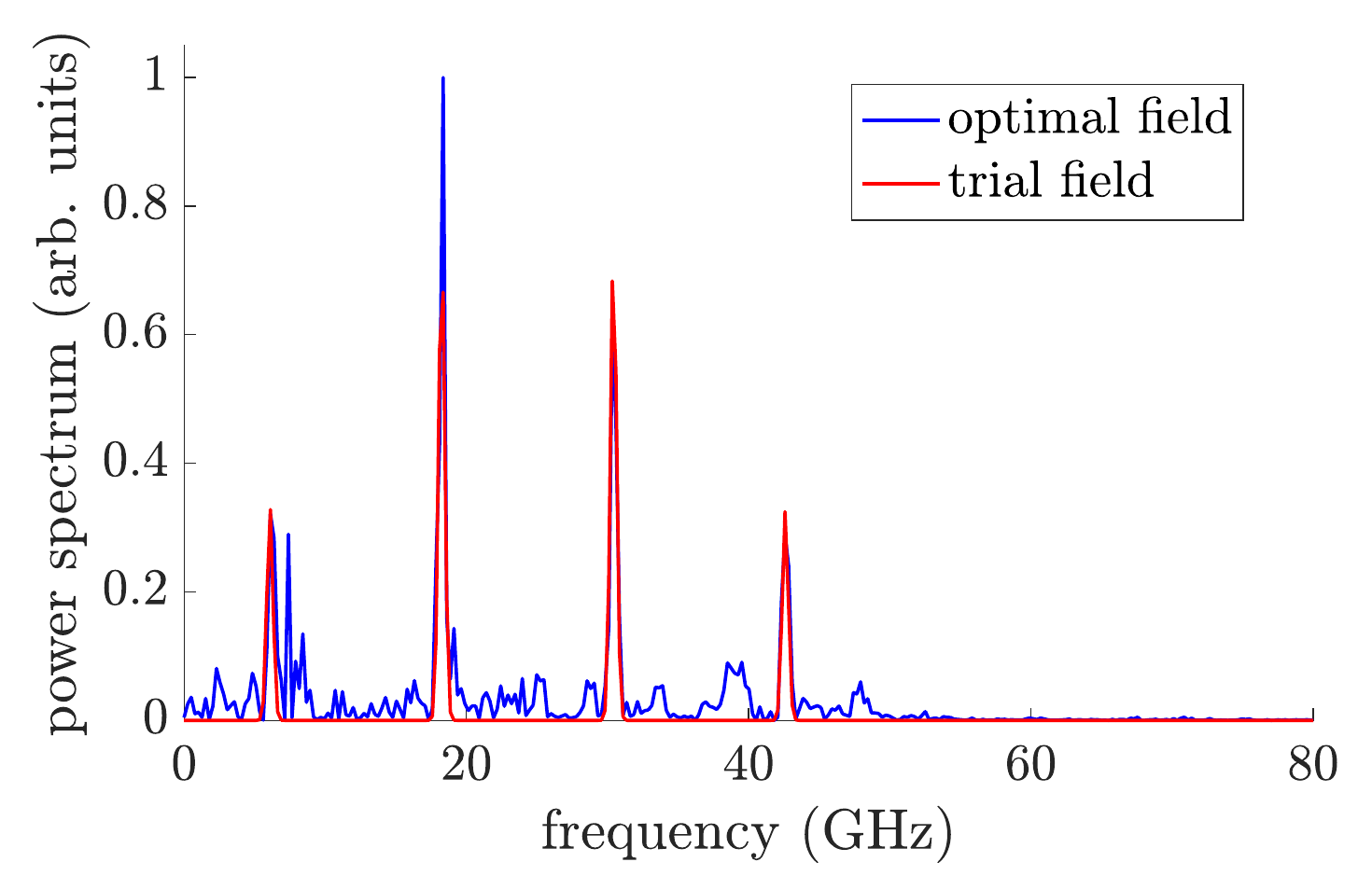}
\caption{Power Spectrum}
\label{fig_2_rotor_ent_freq_field}
\end{subfigure}
\caption{(a) The value of the electric field versus time for the two-rotor entanglement optimization.  We observe that the optimal field (blue) retains the Gaussian envelope of the trial field (red) to a reasonable extent, except that there is increasing amplitude at the start and end points.  (b) The power spectrum (power/frequency vs. frequency) of the electric field for the two-rotor entanglement optimization.  We observe that the optimal field (blue) has five sharp peaks.  Four of them are at the same locations as the initial trial field (red), which correspond to the first four transition frequencies of the field-free non-interacting planar rotor.  The fifth peak is slightly to the right of the first characteristic transition frequency $\omega_1$.}
\end{figure}

In Fig. \ref{fig_2_rotor_ent_temporal_field}, the trial and optimal fields are plotted over time, and in Fig. \ref{fig_2_rotor_ent_freq_field}, the power spectra of these two fields are shown.  It was found that in addition to sharp peaks coinciding with the four characteristic transition frequencies (which were present in the trial field), the optimal field also contained a sharp peak at a frequency lying immediately to the right of the first characteristic transition frequency.

The optimal field $\varepsilon(t)$ yielded an objective of $J^{[2]}_{ent}$ = 0.9560 within the first-order Magnus approximation.  Using this optimal field on a numerically exact model, we calculated an exact objective of $J^{[2]}_{ent}$ = 0.8247.  To further characterize our result, we calculated the von Neumann entropies, resulting in $S_{vN}$ = 2.1131 in the first-order Magnus approximation and $S_{vN}$ = 2.0205 in the exact dynamics.  In comparison, the theoretical maximum possible entropy in the $M=4$ basis is $S_{vN}(\ket{\Psi^{MES}})$ = 2.197 (see Eq.~\ref{eq_maximum_possible_entropy}).  Hence, our optimal control simulation shows that it is possible to optimize to a high objective $J^{[2]}_{ent}$ and drive our system to a state with entropy $S_{vN}$ close to the theoretical maximum.  Importantly, the drop-off in both objective and entropy when using the optimal field found from the approximate dynamics on the exact dynamics is not too large, considering the difficulty of the objective.  We do note, however, that the difference between the approximate and exact objectives is greater in this case than for the two-rotor identical orientation case.

\begin{figure}[h!]
\centering
\includegraphics[width=0.48\textwidth]{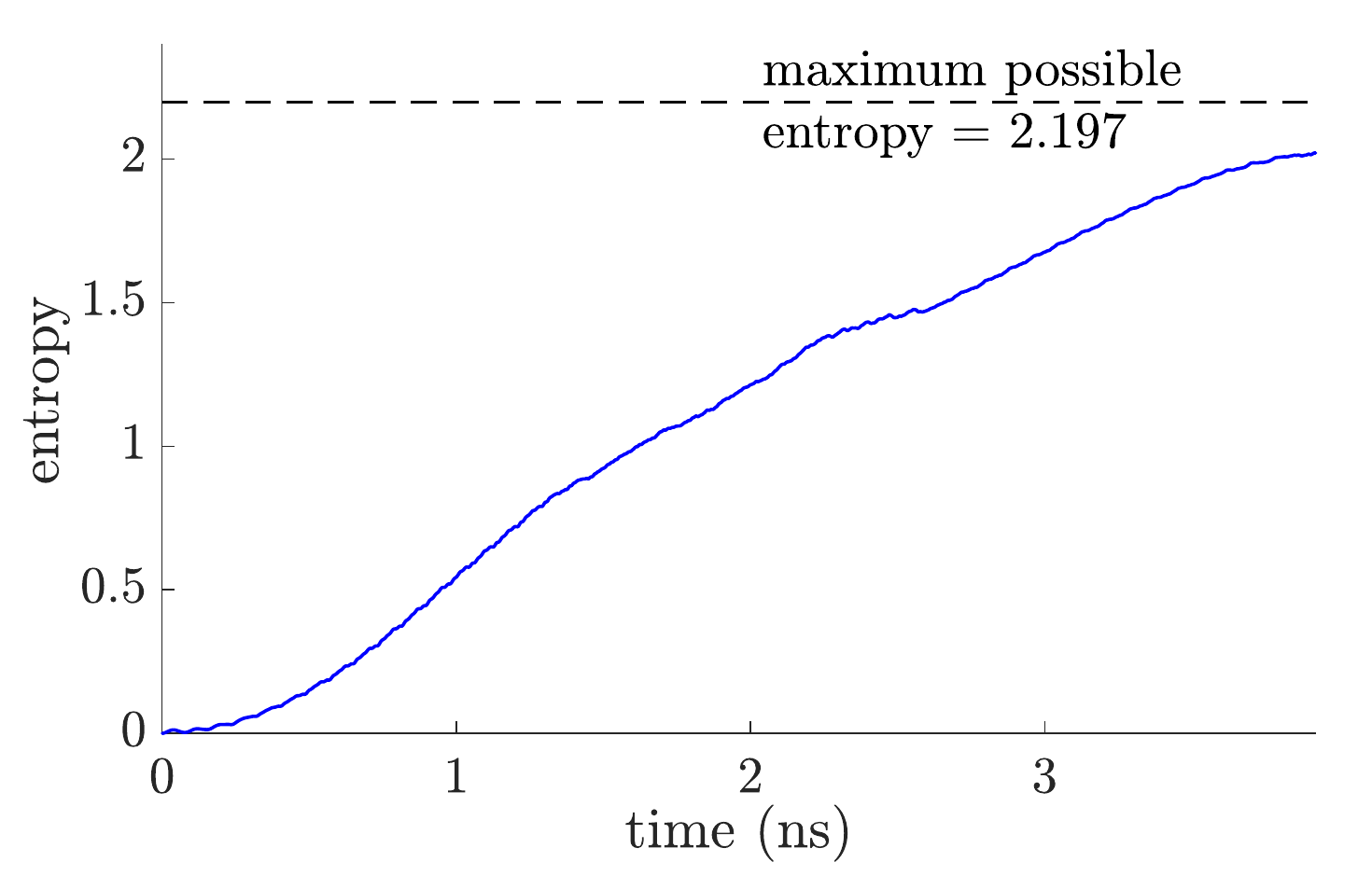}
\caption{Plot of the von Neumann entropy $S_{vN}$ vs. time for exact evolution of two-rotor system in the presence of the optimal field $\varepsilon(t)$, which is the field obtained via optimization of $J^{[2]}_{ent}(T)$ using the first-order Magnus approximation.}
\label{fig_2_rotor_ent_entropy_vs_time}
\end{figure}

In Fig. \ref{fig_2_rotor_ent_entropy_vs_time} the entropy $S_{vN}$ is plotted versus time for evolution under the exact dynamics in the presence $\varepsilon(t)$, which is the optimal field we found using the first-order Magnus approximation.  It was found that the entropy increases steadily, and almost linearly, from zero to optimal value as time progresses.  Our plot in Fig. \ref{fig_2_rotor_ent_entropy_vs_time} is qualitatively similar to the behavior of the system under the optimal field found using the exact solution for entanglement optimization of the two-rotor system presented in \cite{Yu}.

It is important to emphasize that the first-order Magnus approximation within quantum optimal control context can provide a quantitatively accurate mechanism for entanglement, which is of central importance for quantum information applications~\cite{Dolde,Waldherr,Amico,Nielsen}, noting that the zeroth order approximation can only give zero entropy. 

\subsection{Three-Rotor Optimal Control Simulations \label{Subsection_Three_Rotor_Optimal_Control_Simulations}}

\subsubsection{Identical Orientations \label{Subsubsection_Three_Rotor_System_Symmetric_Orientation_Objective}}

Here we perform optimal control calculations to maximize the objective functional $J^{[3]}_{id}(T)$ defined in Eq.~\eqref{eq_objective_functional_three_rotor_idential_orientation} for identical orientation of three OCS rotors (see Fig.~\ref{fig_3_rotor_schematic}).  We choose $R = 6.29$ nm and $T \approx 1.306$ ns, to be approximately consistent with the choice made in \cite{Magann} for Hartree approximation optimization of the same objective Additionally, we chose $M = 5$ for the truncated basis set, $n = 1998$ time steps, and the same trial field that we used in Section~\ref{Subsubsection_Two_Rotor_System_Symmetric_Orientation_Objective}.

The final optimal field $\varepsilon(t)$ for the three-rotor system yielded $\langle \cos\phi_1 \rangle(T) = \langle \cos\phi_3 \rangle(T)$ = 0.9581 and $\langle \cos\phi_2 \rangle(T)$ = 0.9576, within the first-order Magnus approximation.  Applying this optimal field to the exact dynamics, we calculated the exact objectives $\langle \cos\phi_1 \rangle(T) = \langle \cos\phi_3 \rangle(T)$ = 0.9516 and $\langle \cos\phi_2 \rangle(T)$ = 0.9520.  Similar to the two-rotor identical orientations case, we were able to achieve a high value of $J^{[3]}_{id}$, with very little drop-off between approximate and exact dynamics.  In both the approximate and exact dynamics, rotors 1 and 3 and rotor 2 achieve similar orientations at time $T$ (although there is a slight asymmetry).  For reference, applying this same field the exact dynamics in the $M=6$ truncated basis set yields $\langle \cos\phi_1 \rangle(T) = \langle \cos\phi_3 \rangle(T) = 0.9482$ and $\langle \cos\phi_2 \rangle(T) = 0.9477$, indicating that the $M=5$ simulation for the final optimal control field is converged with respect to $M$.

We also examined how the orientations of each rotor evolve versus time for evolution in the exact system in the presence of the optimal field $\varepsilon(t)$ found from the first-order Magnus approximation (not shown).   Although slight differences in the orientation of rotors 1 and 3 and rotor 2 were found throughout the evolution, all three rotors follow roughly the trajectory of time evolution.  Additionally these orientation trajectories had qualitatively similar features as the orientation trajectories of the two-rotor system subject to its corresponding identical orientation optimal field (see Fig. \ref{fig_2_rotor_sym_orientation_vs_time}).  Again, the orientation starts out with a somewhat oscillatory behavior, and acquires its final orientation right at the end of the time range.

\subsubsection{Opposing Orientations \label{Subsubsection_Three_Rotor_System_Anti_Symmetric_Orientation_Objective}}

Here we perform optimal control calculations to maximize the objective functional $J^{[3]}_{opp}(T)$ defined in Eq.~\eqref{eq_objective_functional_three_rotor_opposing_orientation} for opposing orientation of the three-rotor system. We chose $R = 8.5$ nm for the rotor-rotor separation, $T \approx 3.921$ ns for the final time, $n = 5998$ time steps, and the $M=5$ truncated basis set.  The trial field given by Eq. \eqref{eq_trial_field_general_form}, with $F = 3$, $b_0 = 0.25$, $b_1 = 0.15$, $b_2 = 0.6$, and $a_0 = 5.25 \times 10^6$ V/m.

The optimal field $\varepsilon(t)$ yielded the final orientations $\langle \cos\phi_1 \rangle(T) = \langle \cos\phi_3 \rangle(T) = -0.7347$ and $\langle \cos\phi_2 \rangle(T) = 0.8778$, within the first-order Magnus approximation.  Using this optimal field on the exact dynamics, we calculated the exact final orientations $\langle \cos\phi_1 \rangle(T) = \langle \cos\phi_3 \rangle(T) = -0.5888$ and $\langle \cos\phi_2 \rangle(T) = 0.6257$.  We note that compared to the identical orientation optimization $J^{[3]}_{id}$, this optimization achieves a lower value of the objective within the first-order Magnus approximation, and also has a greater drop-off in value between approximate and exact dynamics.  This observation might be explainable by the difficulty of the objective: designing a single global field to create opposing orientations, while relying on interactions that are treated only as a perturbation.  For reference, applying this same field to the exact dynamics in the $M=6$ truncated basis yields $\langle \cos\phi_1 \rangle(T) = \langle \cos\phi_3 \rangle(T) = -0.5877$ and $\langle \cos\phi_2 \rangle(T) = 0.6255$, indicating that the $M= 5$ simulation for this optimal field is converged with respect to $M$.

\begin{figure}[h!]
\centering
\includegraphics[width=0.5\textwidth]{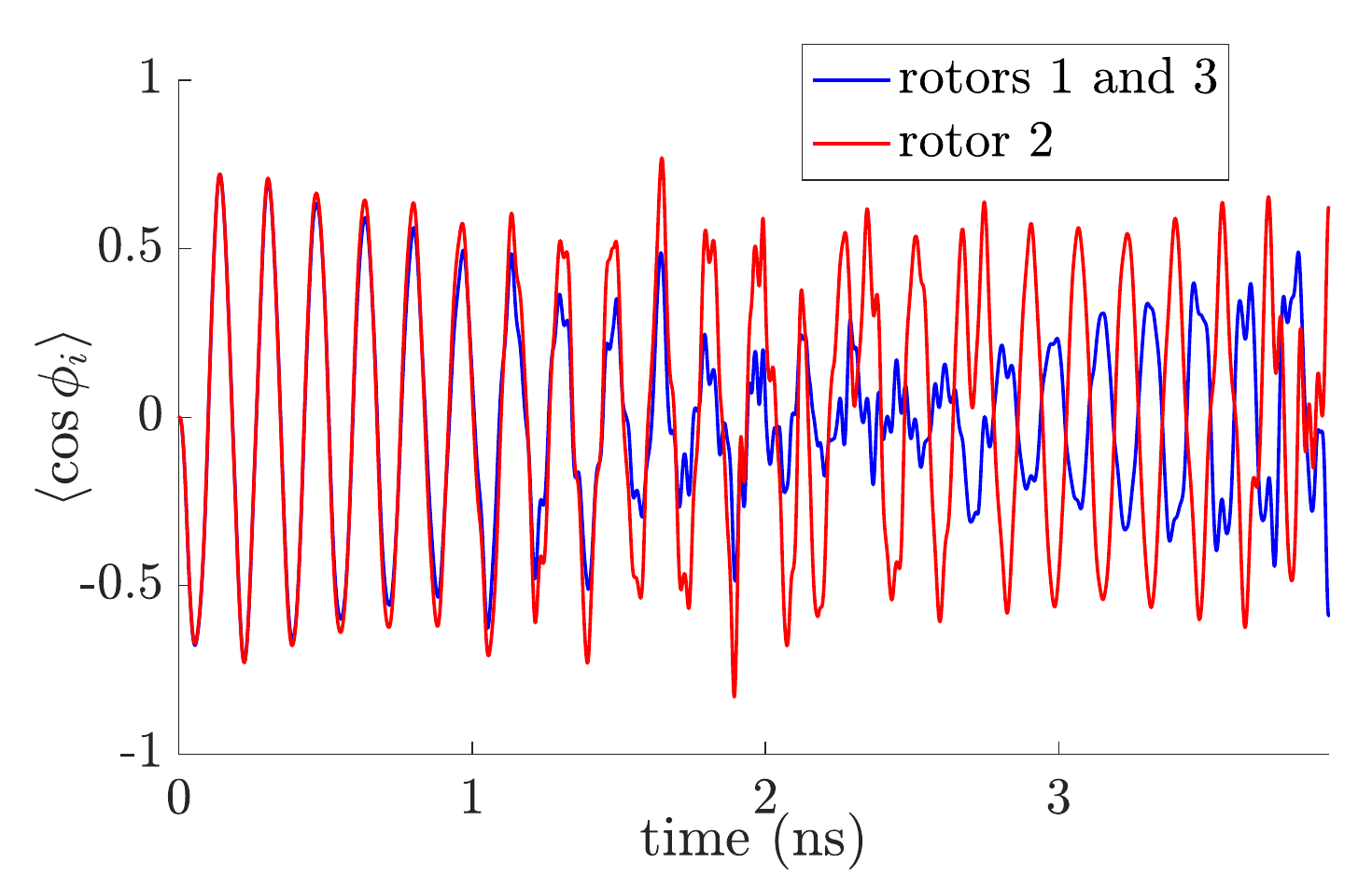}
\caption{Plot of $\langle \cos\phi_1 \rangle(t) = \langle \cos\phi_3 \rangle(t)$ and $\langle \cos\phi_2 \rangle(t)$ vs. time for exact evolution of the three-rotor system in the presence of $\varepsilon(t)$, which is the optimal field obtained from optimizing $J^{[3]}_{opp}(T)$ using the first-order Magnus approximation.  We observe that the oscillations of $\langle \cos\phi_1 \rangle(t) = \langle \cos\phi_3 \rangle(t)$ and $\langle \cos\phi_2 \rangle(t)$ are initially in phase, but are out of phase by about half a period by the end.}
\label{fig_three_rotor_anti_sym_cosine_vs_time}
\end{figure}

In Fig. \ref{fig_three_rotor_anti_sym_cosine_vs_time}, we plot the orientations $\langle \cos\phi_{1,3} \rangle(t)$ and $\langle \cos\phi_2 \rangle(t)$ versus time for exact dynamics in the presence of $\varepsilon(t)$, the optimal field we found using the first-order Magnus approximation.  The orientations of all three rotors oscillate throughout the time period.  The oscillations of $\langle \cos\phi_1 \rangle(t) = \langle \cos\phi_3 \rangle(t)$ and those of $\langle \cos\phi_2 \rangle(t)$ are initially in phase, but are out of phase by roughly half a period by the end of the time period, at which time rotor 2 is oriented in the $+\hat{\textbf{x}}$ direction and rotors 1 and 3 are oriented in the $-\hat{\textbf{x}}$ direction.

This result is of particular interest, because it illustrates that the first-order Magnus approximation can be used to simultaneously control the rotors in non-identical ways, using a global field only, noting that in the zeroth-order approximation all rotors are identical and so non-identical control is impossible.

\section{Conclusions \label{Section_Conclusions}}

In this paper, we have presented a general methodology for simulating the quantum dynamics of systems with multiple weakly-interacting degrees of freedom within the framework of quantum optimal control.  Specifically, we use the first-order Magnus approximation in the interaction picture, treating the couplings between different degrees of freedom as the perturbation.  For optimal control simulations of pairwise coupled quantum systems, the approximation is especially suitable when combined with a gradient-free optimization algorithm, such as the Stochastic Hill Climbing algorithm that we adopt.  For sufficiently many degrees of freedom, we expect the associated computational cost per iteration to be roughly $\frac{1}{n}$ times that of the exact simulation, where $n$ is the number of time steps.

We have demonstrated numerically that this methodology can be used to optimize control fields to drive systems of two and three rotors towards various objectives.  Specifically, for the two-rotor system we provided examples of optimal control of identical orientations and entanglement, and for the three-rotor system we illustrated optimal control of identical orientations and opposing orientations.  In particular, for two-rotor entanglement and three-rotor opposing orientations, the dipole-dipole interactions, despite being treated as the perturbation, are essential for achieving the desired objective.

This work serves as a proof-of-concept demonstration of a new approximate approach to theoretical quantum optimal control of systems with multiple interacting degrees of freedom.  Further research is necessary to characterize the regime of validity, consider different classes of control objectives, and select the most suitable optimization algorithm to be used in conjunction with the quantum dynamics approximation.  Additionally, applying our methodology to large many-body systems would likely require further means for reducing the computational cost.  Towards this end, possible approaches to be considered in future work may include using more efficient numerical methods for matrix operations, exploiting physical symmetries present in many-body systems, or introducing further approximations on top of the first-order Magnus expansion.  In the long term, we hope this work can motivate new approximation methods for many-body quantum optimal control theory, which is an open and challenging area of research.

\section{Acknowledgements \label{Section_Acknowledgements}}

We thank Nicolas Boumal, Nicholas Higham, and Adam Marcus for helpful discussions on numerical methods and computational scaling.  A.B.M. acknowledges support from the DOE CSGF Grant No. DE-FG02-97ER25308. T.S.H. acknowledges support from DOE Grant No. DE-FG02-02ER15344, and H.R. from ARO Grant No. W911NF-16-1-0014.

\bibliographystyle{apsrev}

\begin{thebibliography}{99}
\bibitem{footnote1}  More precisely, it requires computing the product of a matrix exponential and a vector. ($e^A \pmb{v}$).  See e.g. \cite{Al_Mohy} and \cite{Higham}.
\bibitem{footnote2} A multi-step Magnus scheme could be implemented with each step approximating a portion of the time range $[0,T]$.  For example, an evenly-spaced $p$-step propagator is given by $\ket{\Psi(T)} \approx \big( \bigotimes_{i=1}^N U_i(T) \big) \prod_{l=1}^p \bigg( \exp \Big( \Omega_{I,1}\Big(\frac{l}{p}T, \frac{l-1}{p} T \Big) \Big) \bigg) \ket{\Psi(0)}$,
where $\Omega_{I,1}(t_b,t_a) \equiv -\frac{i}{\hbar} \int_{t_a}^{t_b} H_I^{(1)}(t_1)dt_1$.  This $p$-step propagator reduces to the single-step propagator in Eq.~\eqref{eq_first_order_magnus_approximation_applied_to_interacting_system} for $p = 1$ and becomes exact as $p \rightarrow \infty$ (cf. Eqs. \eqref{eq_factored_time_evolution_operator} and \eqref{eq_interaction_time_evolution_operator}).
\bibitem{footnote3} Addtionally, people typically use some factor to account for the costs of additions and other operations.  We will not concern ourselves with that factor here, since we are looking for a rough estimate.~\cite{Moler}
\bibitem{footnote4} We note the caveat that this comparison has assumed that all $D^N$-dimensional computations of the form $e^A \mathbf{v}$ are equally expensive (not just in complexity, but also in actual cost).  However, the matrices and vectors involved in these two simulations are likely different, so the computational costs of each of these computations are not strictly identical.
\end{thebibliography}

\newpage

\appendix
\section{Detailed Numerical Implementation and Cost Analysis for the First-Order Magnus Approximation\label{Appendix_Detailed_Numerical_Procedure}}

Here we present a detailed procedure for numerical implementation of our formulation of the first-order Magnus approximation, as well as an associated computational cost analysis.  We organize this presentation into the three parts outlined in Section \ref{Subsubsection_Fully_Generic_Numerical_Procedure}, corresponding to evaluating Eqs. \eqref{eq_discretized_coupling_free_time_evolution_op}, \eqref{eq_numerical_integration_multi_body_omega_1}, and \eqref{eq_approximate_final_state_numerical_eval} respectively.  For Part II of the numerical implementation, we specialize to the case where the interactions are all pairwise, and discuss why our implementation is computationally cheaper than the naive implementation.

The computational cost analysis is based on our implementation procedure (i.e. we do not claim that there is not a more efficient implementation).  Our computational cost model simply counts the number of complex number multiplications, as stated in Section~\ref{Section_Computational_Cost_Analysis_and_Scaling}.  We use the notation \textit{big} $O$ to denote an upper bound on computational complexity (up to a constant).

For the computational cost analyses in each part, we make use of the following facts and definitions of constants (based on current available matrix algorithms): The computational cost a computing $e^A \mathbf{v}$ given a $d \times d$ matrix $A$ a $d$-dimensional vector $v$ is $C_1 d^{\alpha}$, where $C_1$ and $2 < \alpha \leq 3$ are constants (as stated in Section~\ref{Section_Computational_Cost_Analysis_and_Scaling}).  The computational cost of computing an explicit matrix exponentiation (i.e. computing $e^A$ given $d \times d$ matrix $A$) is upper bounded by $O(d^3)$~\cite{Moler}.  The computational cost of multiplying two $d \times d$ matrices is $C_2 d^{\beta}$, where $C_2$ and $2 < \beta \leq 3$ are constants~\cite{Le_Gall}.  Note that we have not explicitly specified the constants $C_1$, $C_2$, $\alpha$ and $\beta$, because these constants are algorithm and problem-dependent (also, we are concerned with practical algorithms, so the true asymptotic complexity is not always the most relevant).

\subsection{Numerical Implementation Part I \label{Appendix_Subsection_Numerical_Procedure_Part_1}}

For a given $i$, we compute $U_i(k \Delta t)$ for a given $k$ from $U_i((k-1) \Delta t)$ by evaluating Eq. \eqref{eq_discretized_coupling_free_time_evolution_op} as follows:
\begin{enumerate}
\item Form the $D \times D$ matrix $-\frac{i}{\hbar} \Delta t H_i(k \Delta t)$, which has complexity $O(D^2)$.
\item Exponentiate the $D \times D$ matrix $-\frac{i}{\hbar} \Delta t H_i(k \Delta t)$, which has complexity $O(D^3)$ \cite{Moler}.
\item Multiply the two $D \times D$ matrices $
\exp(-\frac{i}{\hbar} \Delta t H_i(k \Delta t))$ and $U_i((k-1) \Delta t)$, which has complexity $O(D^3).$
\end{enumerate}
If there are no symmetries, we need to perform this computation for all $1 \leq i \leq N$ and $1 \leq k \leq n$, so the overall complexity Part I is $O(n N D^3)$.

\subsection{Numerical Implementation Part II \label{Appendix_Subsection_Numerical_Procedure_Part_2}}

In Part II, we compute $\Omega_{I,1}(T)$, which constitutes evaluating Eq.~\eqref{eq_numerical_integration_multi_body_omega_1}.  Specializing to the case where $W$ has the pairwise form given in Eq. \eqref{eq_sum_of_pairwise_interactions}, we can rewrite Eq. \eqref{eq_numerical_integration_multi_body_omega_1} as Eq. \eqref{eq_sum_of_gamma_ij} (cf. Section~\ref{Subsubsection_Exploiting_Separability_to_Reduce_Cost_of_Integration}).  $W$ is a time-independent operator that is given, and from the first part of the procedure we know $U_i(k \Delta t)$ for all $k$ and $i$.  To leverage the separability of $U^{(0)}(t)$ and the pairwise structure of $W$, we proceed as follows:
\begin{enumerate}
\item For a given $i,j$, evaluate $\gamma_{ij}$ within the $D^{2}$-dimensional subspace $\mathcal{\tilde{H}}_i \otimes \mathcal{\tilde{H}}_j$:
\begin{enumerate}
\item At each time step $k$: first compute the tensor product of two $D \times D$ matrices, $U_{i}(k \Delta t) \otimes U_j(k \Delta t)$ (as well as its complex conjugate), which has a complexity of $O(D^4)$.  Then, multiply together the $D^{2} \times D^{2}$ matrices $U_{i}(k \Delta t)^\dagger \otimes U_j(k \Delta t)^\dagger$, $W_{ij}$, and $U_{i}(k \Delta t) \otimes U_j(k \Delta t)$, which has a complexity of $O(D^6)$.
\item Add together the resulting matrices $\big( U_{i}(k \Delta t)^\dagger \otimes U_j(k \Delta t)^\dagger \big) W_{ij} \big( U_{i}(k \Delta t) \otimes U_j(k \Delta t) \big)$ from each time $k$.
\item Multiply the resulting $D^2 \times D^2$ matrix from part (b) by the constant $-\frac{i}{\hbar} \Delta t$, which has a complexity of $O(D^2)$.
\end{enumerate}
\item Convert each $\gamma_{ij}$ from its representation in the subspace $\mathcal{\tilde{H}}_i \otimes \mathcal{\tilde{H}}_j$ to its representation in the $D^N$-dimensional state space $\mathcal{\tilde{H}}$.  (This does not involve any multiplications.)
\item Finally, in the $D^N$-dimensional state space $\mathcal{\tilde{H}}$, add together all of the $\gamma_{ij}$.
\end{enumerate}
In the worst case, there are a total of $\frac{N(N-1)}{2}$ distinct $\gamma_{ij}$, and so the computational of the first step is $O(n N^2 D^6)$.  The second and third steps do not involve any complex number multiplications.  Therefore, the overall computational cost of this implementation of Part II has an upper bound of $O(n N^2 D^6)$. 

A naive implementation of Part II, without taking advantage of the pairwise structure of $W$, would be to directly evaluate Eq.~\eqref{eq_numerical_integration_multi_body_omega_1}.  Without counting the detailed computational costs, we can see that a direct evaluation would require multiplying the $D^N \times D^N$ matrices $\big(\bigotimes_{i=1}^N U_i(k \Delta t)^\dagger$\big), W, and $\big(\bigotimes_{i'=1}^N U_{i'}(k \Delta t)\big)$ at every time step $k = 1,2,...,n$.  Therefore, the naive method has a computational cost of at least $2 C_2 n D^{\beta N}$.  So we can see that the method leveraging separability and pairwise interactions reduces the computational cost of Part II from exponential with respect to $N$ to polynomial with respect to $N$.

\subsection{Numerical Implementation Part III \label{Appendix_Subsection_Numerical_Procedure_Part_3}}

In Part III, we compute $\ket{\Psi(T)}$ by evaluating Eq. \eqref{eq_approximate_final_state_numerical_eval} as follows:
\begin{enumerate}
\item Compute $\exp(\Omega_{I,1}(T)) \ket{\Psi(0)}$.  This is one $D^N$-dimensional computation of the form $e^A \mathbf{v}$, and so the computational cost of this step is $C_1 D^{\alpha N}$.
\item Compute the tensor product of $N$ $D \times D$ matrices, $\bigotimes_{i = 1}^N U_i(T)$ from the $U_i(n \Delta t)$, which has a cost of $D^{2N} + O(D^{N + 1})$.
\item Multiply together the $D^N \times D^N$ matrix $\bigotimes_{i = 1}^N U_i(T)$ from step 2 and the $D^N$-dimensional vector $\exp(\Omega_{I,1}(T)) \ket{\Psi(0)}$ from step 1.  This has a cost of $D^{2N}$.
\end{enumerate}
Therefore, the overall computational cost of Part III is $C_1 D^{\alpha N} + 2 D^{2N} + O(D^{N+1})$.

\section{Second-Order Magnus Approximation \label{Appendix_Second_Order_Magnus_Approximation}}

Here we consider the second-order Magnus approximation in the interaction picture for simulating an interacting quantum system.  We present the mathematical formulation in Appendix~\ref{Appendix_Subsection_MA2_Formulation} and the numerical implementation and associated computational cost analysis in Appendix~\ref{Appendix_Subsection_MA2_Numerical_Implementation}.  Note that we do not implement the second-order Magnus approximation in any of our numerical simulations.

\subsection{Formulation \label{Appendix_Subsection_MA2_Formulation}}

We use the same choices of $H^{(1)}$ and $H^{(0)}$ as in Section~\ref{Subsection_Approximate_Simulation_of_Weakly_Coupled_Multi_Dimensional_Quantum_Dynamics} (i.e. Eqs. \eqref{eq_first_order_Hamiltonian} and \eqref{eq_zeroth_order_Hamiltonian} respectively).  The second-order Magnus approximation wavefunction at final time $t=T$ is given by:
\begin{equation}
\ket{\Psi(T)} \approx \big( \bigotimes_{i=1}^N U_i(T) \big) \exp(\Omega_{I,1}(T) + \Omega_{I,2}(T)) \ket{\Psi(0)}, \label{eq_second_order_Magnus_approximation_interacting_systems}
\end{equation}
where the $U_i(t)$ are given by Eq.~\eqref{eq_coupling_free_time_evolution_operator}, $\Omega_{I,1}(T)$ is given by Eq.~\eqref{eq_omega_I_1_for_interacting_systems}, and $\Omega_{I,2}(T)$ is given by
\begin{equation}
\begin{aligned}
\Omega_{I,2}(T) &= -\frac{1}{2 \hbar^2} \int_0^{T} dt_1 \int_0^{t_1} dt_2 \Big[\big(\bigotimes_{i=1}^N U_{i}(t_1)^{\dagger}) W \\
&\times \big(\bigotimes_{i'=1}^N U_{i'}(t_1)\big), \big(\bigotimes_{i''=1}^N U_{i''}(t_2)^{\dagger}) W \big(\bigotimes_{i'''=1}^N U_{i'''}(t_2)\big)\Big], 
\end{aligned}
\end{equation}
cf. Eq.~\eqref{eq_Omega_2}.  As with the first-order Magnus approximation, Eq.~\eqref{eq_second_order_Magnus_approximation_interacting_systems} computes the final state without computing the state at intermediate times.

\subsection{Numerical Implementation and Computational Cost Analysis \label{Appendix_Subsection_MA2_Numerical_Implementation}}

In order to numerically implement the second-order Magnus approximation, we can simply modify the three-part procedure for implementing the first-order Magnus approximation (outlined in Section~\ref{Subsection_Numerical_Implementation_of_Our_General_Methodology} and detailed in Appendix~\ref{Appendix_Detailed_Numerical_Procedure}).  Part I is unchanged.  For Part II, in addition to computing $\Omega_{I,1}(T)$ as was done in the first-order Magnus approximation, we also need to compute $\Omega_{I,2}(T)$ and then add it to $\Omega_{I,2}(T)$.  For Part III, we simply replace $\Omega_{I,1}(T)$ with $\Omega_{I,1}(T) + \Omega_{I,2}(T)$ and evaluate in the same way.  We will use the same computational cost model as before (i.e. counting the number of complex number multiplications and ignoring other operations), and so the computational cost of the second-order Magnus approximation will be equal to the cost of the first-order Magnus approximation plus the cost of computing $\Omega_{I,2}(T)$.  Below, we detail the steps for computing $\Omega_{I,2}(T)$, assuming that $W$ takes the pairwise form given in Eq.~\eqref{eq_sum_of_pairwise_interactions}.

After discretization into $n$ uniform time steps, the double integral for $\Omega_{I,2}(T)$ is given by
\begin{equation}
\begin{aligned}
\Omega_{I,2}(T) &= -\frac{1}{2 \hbar^2} \sum_{k=1}^n \Delta t \sum_{k' = 1}^k \Delta t \Big[\big(\bigotimes_{i=1}^N U_{i}(k \Delta t)^{\dagger}) W \\
&\times\big(\bigotimes_{i'=1}^N U_{i'}(k \Delta t)\big), \big(\bigotimes_{i''=1}^N U_{i''}(k' \Delta t)^{\dagger}) W \\
&\times\big(\bigotimes_{i'''=1}^N U_{i'''}(k' \Delta t)\big)\Big]. \label{eq_Omega_I_2_numerical_integral}
\end{aligned}
\end{equation}
where we have invoked the $n$-point rectangular quadrature for the double integration with respect to time $t$ over the interval $[0,T]$.  Substituting Eq.~\eqref{eq_sum_of_pairwise_interactions} into Eq.~\eqref{eq_Omega_I_2_numerical_integral}, we have
\begin{equation}
\Omega_{I,2}(T) = \sum_{1 \leq i < j \leq N} \sum_{1 \leq i' < j' \leq N} \kappa_{iji'j'}, \label{eq_Omega_I_2_sum_of_kappa_iji'j'}
\end{equation}
where
\begin{equation}
\begin{aligned}
\kappa_{iji'j'} & \equiv -\frac{(\Delta t)^2}{2 \hbar^2} \sum_{k=1}^n \sum_{k' = 1}^k \Big[\big(\bigotimes_{i''=1}^N U_{i''}(k \Delta t)^{\dagger}) W_{ij} \\
&\times \big(\bigotimes_{i'''=1}^N U_{i'''}(k \Delta t)\big),\big(\bigotimes_{i''''=1}^N U_{i''''}(k' \Delta t)^{\dagger}\big) W_{i'j'}\\
&\times\big(\bigotimes_{i'''''=1}^N U_{i'''''}(k' \Delta t)\big)\Big]. \label{eq_kappa_iji'j'_general_form}
\end{aligned}
\end{equation}
Using the fact that $W_{ij}$ only acts on the $i$-th and $j$-th degrees of freedom, Eq.~\eqref{eq_kappa_iji'j'_general_form} becomes
\begin{multline}
\kappa_{iji'j'} = -\frac{(\Delta t)^2}{2 \hbar^2} \sum_{k=1}^n \sum_{k' = 1}^k \Big[\big(U_{i}(k\Delta t)^{\dagger} \otimes U_{j}(k\Delta t)^{\dagger} \big) W_{ij}\\ \times\big(U_{i}(k\Delta t) \otimes U_{j}(k\Delta t) \big),
 \big(U_{i'}(k'\Delta t)^{\dagger} \otimes U_{j'}(k'\Delta t)^{\dagger} \big) W_{i'j'}\\ \times\big(U_{i'}(k'\Delta t) \otimes U_{j'}(k'\Delta t) \big)\Big].
\end{multline}
Note that for all $i$, $j$, and $k$, the term $\big(U_{i}(k\Delta t)^{\dagger} \otimes U_{j}(k\Delta t)^{\dagger} \big) W_{ij} \big(U_{i}(k\Delta t) \otimes U_{j}(k\Delta t) \big)$ is already computed as an intermediate step when computing $\Omega_{I,1}(T)$ (see Appendix \ref{Appendix_Subsection_Numerical_Procedure_Part_2}).  $\kappa_{iji'j'} = 0$ for $\{i,j\} \cap \{i',j'\} = \emptyset$, so Eq. \eqref{eq_Omega_I_2_sum_of_kappa_iji'j'} becomes
\begin{equation}
\Omega_{I,2}(T) =  \sum_{1 \leq i < j \leq N} \sum_{\substack{1 \leq i' < j' \leq N \\ i'=i, \, i'=j, \, j'=i, \, or \, j'=j}} \kappa_{iji'j'}. \label{eq_Omega_I_2_sum_of_five_types_of_terms}
\end{equation}

Now, we can compute $\Omega_{I,2}(T)$ by evaluating Eq. \eqref{eq_Omega_I_2_sum_of_five_types_of_terms} with the following numerical procedure (note that the following is not intended to be the most efficient procedure, and is primarily intended to show in a straightforward way that the computation can be done in polynomial cost with respect to $N$):
\begin{enumerate}
\item Compute each $\kappa_{iji'j'}$ in the sum in Eq.~\eqref{eq_Omega_I_2_sum_of_five_types_of_terms} that $i'=i$ and $j'=j$ (i.e. terms of the form $\kappa_{ijij}$) as follows:
\begin{enumerate}
\item At each combination of $k$ and $k'$, within the subspace $\mathcal{\tilde{H}}_i \otimes \mathcal{\tilde{H}}_j$, compute the commutator \\ $\Big[\big(U_{i}(k\Delta t)^{\dagger} \otimes U_{j}(k\Delta t)^{\dagger} \big) W_{ij} \big(U_{i}(k\Delta t) \otimes U_{j}(k\Delta t) \big),
\\ \big(U_{i}(k'\Delta t)^{\dagger} \otimes U_{j}(k'\Delta t)^{\dagger} \big) W_{ij} \big(U_{i}(k'\Delta t) \otimes U_{j}(k'\Delta t) \big)\Big]$.  This involves two $D^2 \times D^2$ matrix multiplications, which has a complexity of $O(D^{6})$.
\item Add together the result from step (a) at every combination of $k$ and $k'$.
\item Multiply the resulting $D^2 \times D^2$ matrix from part (b) by the constant $-\frac{1}{2 \hbar^2}$ to get $\kappa_{ijij}$, which has a complexity of $O(D^4)$.
\end{enumerate}
\item Compute each term $\kappa_{iji'j'}$ in the sum in Eq.~\eqref{eq_Omega_I_2_sum_of_five_types_of_terms} that were not already computed in Step 1 (i.e. terms of the form $\kappa_{ijij'}$, $\kappa_{iji'i}$, $\kappa_{ijjj'}$, and $\kappa_{iji'j}$) as follows:
\begin{enumerate}
\item At each combination of $k$ and $k'$, convert the matrices $\big(U_{i}(k\Delta t)^{\dagger} \otimes U_{j}(k\Delta t)^{\dagger} \big) W_{ij} \big(U_{i}(k\Delta t) \otimes U_{j}(k\Delta t) \big)$ and $\big(U_{i'}(k'\Delta t)^{\dagger} \otimes U_{j'}(k'\Delta t)^{\dagger} \big) W_{i'j'} \big(U_{i'}(k'\Delta t) \otimes U_{j'}(k'\Delta t) \big)$ (where it is understood, for example, that for terms of the form $\kappa_{ijjj'}$ that $i'=j$ in the preceding matrix) into the corresponding subspace of three degrees of freedom (e.g. for terms of the form $\kappa_{ijjj'}$, this would be $\mathcal{\tilde{H}}_i \otimes \mathcal{\tilde{H}}_j \otimes \mathcal{\tilde{H}}_{j'}$), which has negligible cost.  Then compute the commutator \\ $\Big[\big(U_{i}(k\Delta t)^{\dagger} \otimes U_{j}(k\Delta t)^{\dagger} \big) W_{ij} \big(U_{i}(k\Delta t) \otimes U_{j}(k\Delta t) \big),
\\ \big(U_{i'}(k'\Delta t)^{\dagger} \otimes U_{j'}(k'\Delta t)^{\dagger} \big) W_{i'j'} \big(U_{i'}(k'\Delta t) \otimes U_{j'}(k'\Delta t) \big)\Big]$.  This involves two $D^3 \times D^3$ matrix multiplications, which has a complexity of $O(D^9)$.
\item Add together the result from step (a) at every combination of $k$ and $k'$.
\item Multiply the resulting $D^3 \times D^3$  matrix from part (b) by the constant $-\frac{1}{2 \hbar^2}$ to get $\kappa_{iji'j}$, which has a complexity of $O(D^6)$.
\end{enumerate}
\item Convert all of the terms $\kappa_{iji'j'}$ in the sum in Eq.~\eqref{eq_Omega_I_2_sum_of_five_types_of_terms} (which we have computed in Steps 2 and 3) from the subspace of two or three degrees of freedom to the full state space $\mathcal{\tilde{H}}$ (which involves no multiplications), and then add them together. 
\end{enumerate}
Therefore, the total computational complexity of computing $\Omega_{I,2}(T)$ will be the complexity of computing each $\kappa_{ijij}$, $\kappa_{ijij'}$, $\kappa_{iji'i}$, $\kappa_{ijjj'}$, and $\kappa_{iji'j}$.  Since there are $O(n^2)$ combinations of $k$ and $k'$, the complexity of computing each $\kappa_{ijij}$ will be $O(n^2 D^6)$, and the complexity of computing each $\kappa_{ijij'}$, $\kappa_{iji'i}$, $\kappa_{ijjj'}$, and $\kappa_{iji'j}$ will be $O(n^2 D^9)$.  There are at most $O(N^2)$ terms of the form $\kappa_{ijij}$, and at most $O(N^3)$ terms of the forms $\kappa_{ijij'}$, $\kappa_{iji'i}$, $\kappa_{ijjj'}$, and $\kappa_{iji'j}$.  Therefore, the total computational complexity of computing $\Omega_{I,2}(T)$ will be upper bounded by $O(n^2 N^3 D^9)$.

Therefore, the total computational cost of one full simulation with the second-order Magnus approximation will be $C_1 D^{\alpha N} + 2 D^{2N} + O(D^{N+1}) + O(n^2 N^3 D^9)$.

\section{Accuracy vs. Separation of the First-Order Magnus Approximation for Two and Three-Rotor Simulations \label{Appendix_Accuracy_vs_Separation_of_First_Order_Magnus_Approximation_for_Two_and_Three_Rotor_Simulations}}

\begin{figure}[h!]
\centering
\begin{subfigure}[b]{0.45\textwidth}
\includegraphics[width=\textwidth]{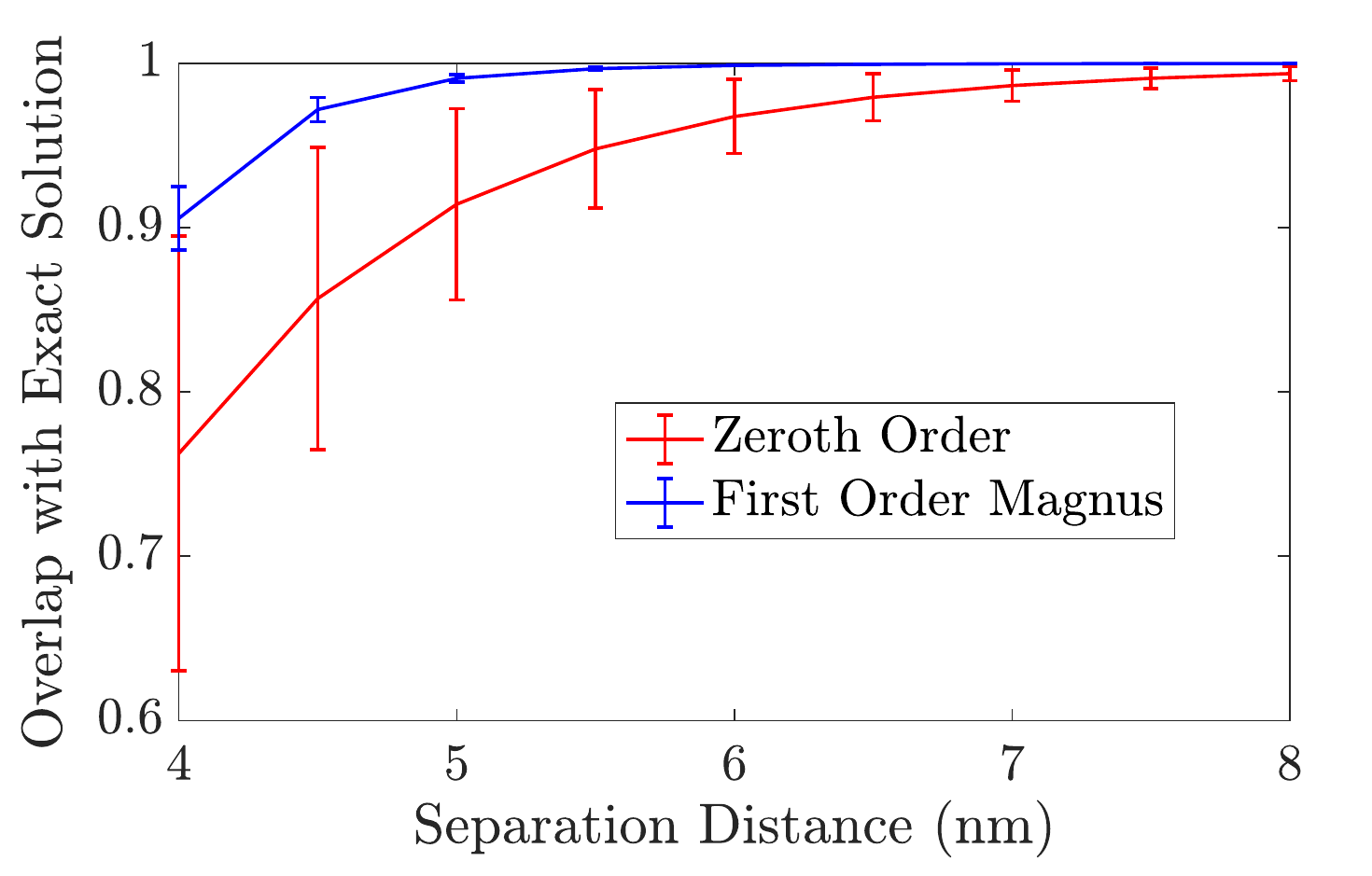}
\caption{Two-Rotor Overlap vs. Separation Distance} \label{fig_2_rotor_overlap}
\end{subfigure}
\begin{subfigure}[b]{0.45\textwidth}
\includegraphics[width=\textwidth]
{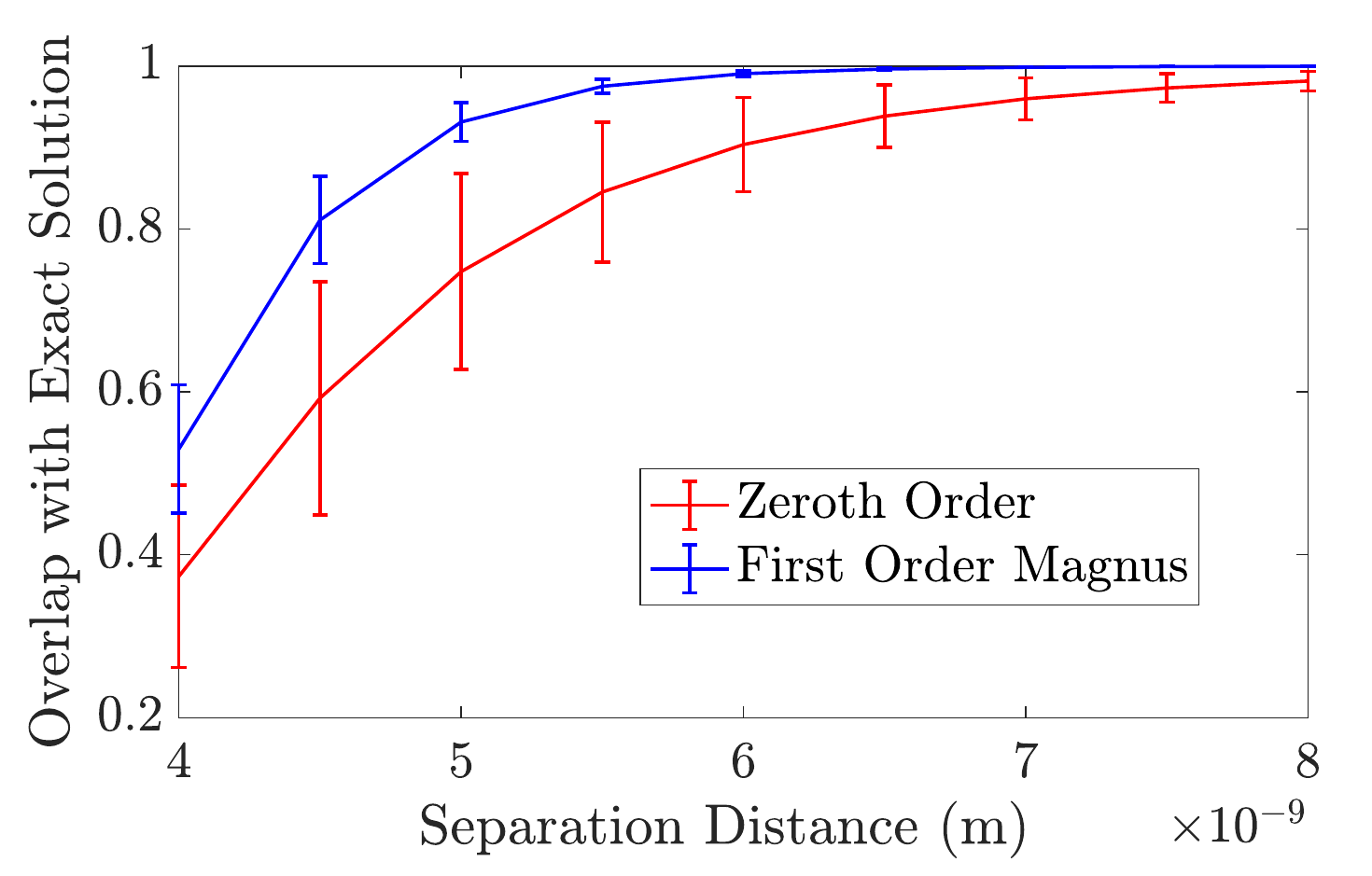}
\caption{Three-Rotor Overlap vs. Separation Distance}
\label{fig_3_rotor_overlap}
\end{subfigure}
\caption{Overlap $|\bra{\Psi_{approx}(T)}\ket{\Psi_{exact}(T)}|$ vs. separation.  In both (a) and (b), the blue curve corresponds to $\ket{\Psi_{approx}}$ being the first order Magnus expansion, and the red curve corresponds to $\ket{\Psi_{approx}}$ being the zeroth order wavefunction, which neglects interactions between rotors.  The values represent the average value from a collection of five electric fields, and the error bars are one standard deviation. In (a), we plot the results for the two-rotor system with $\theta_{12}=\frac{\pi}{2}$.  In (b), we plot the results for the three-rotor system arranged in an equilateral triangle geometry.  In both plots, we observe that the first order Magnus expansion improves the quality of the approximation over the zeroth order approximation.  We also observe in both plots that the quality of both the zeroth order and first order approximations get worse with decreasing separation.}
\end{figure}

The purpose of this section is to investigate numerically how separation $R$ affects the accuracy of the first-order Magnus approximation.  Moreover, we want to understand the extent to which the first-order Magnus approximation improves upon the zeroth order approximation.  We specialize to the geometries and initial states stated in Sections \ref{Subsubsection_Model_Hamiltonian_for_Two_Coupled_Molecular_Rotors} and \ref{Subsubsection_Model_Hamiltonian_for_Three_Coupled_Molecular_Rotors} for the two and three-rotor systems respectively.

We compare the zeroth order simulation, first order Magnus simulation, and exact simulations at different separations for a collection of arbitrary (not optimized) test fields.  Note that we implement the zeroth order approximation using small time step operators as well, cf. Eq. \eqref{eq_short_timestep_propagator}, i.e.
\begin{equation}
\ket{\Psi^{(0)}(T)} = \bigotimes_{i = 1}^N \Big( U_i(T) \ket{\psi_i^{(0)}(0)} \Big),
\end{equation}
where $\ket{\psi_i^{(0)}(t)}$ is the wavefunction of the $i$-th degree of freedom in the zeroth order approximation and  $\ket{\Psi^{(0)}(t)}$ is the full zeroth order wavefunction.

In this Section, we use $T \approx 1.306$ ns, $M = 4$, and $n = 999$.  We randomly generated five different fields of the form:
\begin{equation}
\epsilon_j(t) = a_0 \sum_{m=0}^{3} \tilde{b}_m \cos(\omega_m t + \tilde{\delta}_m),
\end{equation}
for $j=1,2,...,5$, where $a_0 = 5 \times 10^6 \frac{N}{C}$, $\tilde{b}_m$ and $\tilde{\delta}_m$ are random numbers generated between (0,1) and (0,2$\pi$) respectively, and $\omega_m = \frac{B (2m + 1)}{\hbar}$ is the field-free, interaction-free, individual rotor transition frequency from $\ket{m}$ to $\ket{m+1}$.  For both the two and three-rotor systems, at each separation $R$ and for each field, we simulated the dynamics using the zeroth-order approximation $\ket{\Psi^{(0)}}$, first-order Magnus approximation $\ket{\Psi_{Magnus}}$, and exact dynamics $\ket{\Psi_{exact}}$.  We then computed $|\bra{\Psi^{(0)}(T)}\ket{\Psi_{exact}(T)}|$ and $|\bra{\Psi_{magnus}(T)}\ket{\Psi_{exact}(T)}|$, which are plotted versus separation in Figs. \ref{fig_2_rotor_overlap} and \ref{fig_3_rotor_overlap} for the two and three-rotor systems respectively.

From these figures, we observe that the first order Magnus approximation improves clearly upon the zeroth order approximation at all separations, and that as expected both approximations are better at larger separations (since the perturbation strength scales as $\frac{1}{R^3}$).  Additionally, the first-order approximation has a much smaller error bar than the zeroth order approximation at all separations.  We also observe that for the same separation, both the first-order Magnus and zeroth order approximations have lower overlap in the three-rotor case than in the two-rotor case (in the three-rotor configuration, each rotor feels coupling from two rotors, and so the total magnitude of the interaction is greater).  Finally, we point out that although this cannot be seen directly from the plots of the average overlaps for all fields, for a given field, both the Magnus approximation and the zeroth order approximation are strictly better at increasing separation.  Note that the fields we tested are not by any means exhaustive, and the quality of approximation likely depends on other factors that we did not vary, such as field amplitude and length of time.  These tests are intended only to give a ballpark estimate.

\section{Proof of Three-Rotor Symmetries \label{Appendix_Three_Rotor_Symmetries}}

In Appendix \ref{Appendix_Three_Rotor_Symmetries}, we present an analytical proof of Eqs.~\eqref{eq_cosphi1_cosphi_3_relationship}, \eqref{eq_sinphi1_sinphi3_relationship}, and \eqref{eq_sinphi2_relationship}, by first proving a more general result.

\noindent\underline{\textbf{Theorem 1}}

Let $H(\phi_1,\phi_2,\phi_3;t)$ be an arbitrary Hamiltonian that acts on three degrees of freedom with identical Hilbert spaces $V$ (i.e. $H: \mathcal{V} \otimes \mathcal{V} \otimes \mathcal{V} \rightarrow \mathcal{V} \otimes \mathcal{V} \otimes \mathcal{V}$).  Assume that the system (and thus all observables) are invariant under the transformations $\phi_i \rightarrow \phi_i + 2 \pi$, for $i=1,2,3$.  Suppose
\begin{equation}
H(\phi_1 \rightarrow \phi_3, \phi_2, \phi_3 \rightarrow \phi_1; t) = H(-\phi_1,-\phi_2,-\phi_3;t), \label{eq_condition_1_of_theorem_1}
\end{equation}
and let $\Psi_0(\phi_1,\phi_2,\phi_3)$ be an arbitrary state that satisfies
\begin{equation}
\Psi_0(\phi_1 \rightarrow \phi_3, \phi_2, \phi_3 \rightarrow \phi_1) = \Psi_0(-\phi_1,-\phi_2,-\phi_3). \label{eq_condition_2_of_theorem_1}
\end{equation}
Let $\Psi(\phi_1,\phi_2,\phi_3;t)$ be the solution to the Schr\"odinger equation for Hamiltonian $H(\phi_1,\phi_2,\phi_3;t)$, subject to
\begin{equation}
\Psi(\phi_1,\phi_2,\phi_3;t=0) = \Psi_0(\phi_1,\phi_2,\phi_3).
\end{equation}
Then for any observable $A(\phi_i)$ that acts on a single degree of freedom, and for all times $t$,
\begin{equation}
\langle A(\phi_1) \rangle(t) = \langle A(-\phi_3) \rangle(t) \label{eq_theorem_1_part_1}
\end{equation}
and
\begin{equation}
\langle A(\phi_2) \rangle(t) = \langle A(-\phi_2) \rangle(t), \label{eq_theorem_1_part_2}
\end{equation}
where 
\begin{equation}
    \langle A(\phi_j) \rangle (t) \equiv \bra{\Psi(\phi_1,\phi_2,\phi_3;t)}A(\phi_j)\ket{\Psi(\phi_1,\phi_2,\phi_3;t)}.
\end{equation}

\noindent \underline{\textbf{Proof of Theorem 1}}

Define $\tilde{H}(\phi_1,\phi_2,\phi_3;t)$ as
\begin{equation}
\begin{aligned}
\tilde{H}(\phi_1,\phi_2,\phi_3;t) &\equiv H(\phi_1 \rightarrow \phi_3, \phi_2, \phi_3 \rightarrow \phi_1; t) \\
&= H(-\phi_1,-\phi_2,-\phi_3;t). \label{eq_H_prime_definition}
\end{aligned}
\end{equation}
Define $\tilde{\Psi}_0(\phi_1,\phi_2,\phi_3)$ as 
\begin{equation}
\begin{aligned}
\tilde{\Psi}_0(\phi_1,\phi_2,\phi_3) & \equiv \Psi_0(\phi_1 \rightarrow \phi_3, \phi_2, \phi_3 \rightarrow \phi_1) \\
&= \Psi_0(-\phi_1,-\phi_2,-\phi_3). \label{eq_Psi_prime_definition}
\end{aligned}
\end{equation}
Define $\tilde{\Psi}_0(\phi_1,\phi_2,\phi_3;t)$ as the solution to the Schr\"odinger equation for $\tilde{H}(\phi_1,\phi_2,\phi_3;t)$, subject to
\begin{equation}
\tilde{\Psi}(\phi_1,\phi_2,\phi_3;t=0) = \tilde{\Psi}_0(\phi_1,\phi_2,\phi_3).
\end{equation}
Eqs. \eqref{eq_H_prime_definition} and \eqref{eq_Psi_prime_definition} indicate that the system governed by $H$ and $\Psi_0$ and the system governed by $\tilde{H}$ and $\tilde{\Psi}_0$ can be related by simply relabeling the indices $\phi_1 \rightarrow \phi_3$ and $\phi_3 \rightarrow \phi_1$.  This implies
\begin{equation}
\tilde{\Psi}(\phi_1,\phi_2,\phi_3; t) = \Psi(\phi_1 \rightarrow \phi_3, \phi_2, \phi_3 \rightarrow \phi_1; t),
\end{equation}
and so
\begin{equation}
\begin{aligned}
\bra{\tilde{\Psi}(\phi_1,\phi_2,\phi_3;t)}&A(\phi_3)\ket{\tilde{\Psi}(\phi_1,\phi_2,\phi_3;t)} \\
&= \bra{\Psi(\phi_1,\phi_2,\phi_3;t)}A(\phi_1)\ket{\Psi(\phi_1,\phi_2,\phi_3;t)} \label{eq_EV_relationship_between_Psi_prime_and_Psi_for_phi1_andphi3}
\end{aligned}
\end{equation}
and
\begin{equation}
\begin{aligned}
\bra{\tilde{\Psi}(\phi_1,\phi_2,\phi_3;t)}&A(\phi_2)\ket{\tilde{\Psi}(\phi_1,\phi_2,\phi_3;t)} \\
&= \bra{\Psi(\phi_1,\phi_2,\phi_3;t)}A(\phi_2)\ket{\Psi(\phi_1,\phi_2,\phi_3;t)}. \label{eq_EV_relationship_between_Psi_prime_and_Psi_for_phi2}
\end{aligned}
\end{equation}
Moreover, we know from taking the change of variables $\phi_1 \rightarrow -\phi_1, \phi_2 \rightarrow -\phi_2, \phi_3 \rightarrow -\phi_3$, that $\Psi(-\phi_1,-\phi_2,-\phi_3;t)$ will be the solution to the Schr\"odinger equation with Hamiltonian $H(-\phi_1,-\phi_2,-\phi_3;t)$ and initial state $\Psi_0(-\phi_1,-\phi_2,-\phi_3)$.  From Eqs. \eqref{eq_H_prime_definition} and \eqref{eq_Psi_prime_definition}, $H(-\phi_1,-\phi_2,-\phi_3;t) = \tilde{H}(\phi_1,\phi_2,\phi_3;t)$ and $\Psi_0(-\phi_1,-\phi_2,-\phi_3) = \tilde{\Psi}_0(\phi_1,\phi_2,\phi_3)$.  Therefore, since the solution to the Schr\"odinger equation is unique,
\begin{equation}
\tilde{\Psi}(\phi_1,\phi_2,\phi_3;t) = \Psi(-\phi_1,-\phi_2,-\phi_3;t),
\end{equation}
and so for $i$ = 1, 2, or 3, by direct computation, we have
\begin{equation}
\begin{aligned}
\bra{\tilde{\Psi}(\phi_1,\phi_2,\phi_3;t)}&A(\phi_i)\ket{\tilde{\Psi}(\phi_1,\phi_2,\phi_3;t)} \\
&= \bra{\Psi(\phi_1,\phi_2,\phi_3;t)}A(-\phi_i)\ket{\Psi(\phi_1,\phi_2,\phi_3;t)}. \label{eq_EV_relationship_between_Psi_prime_and_Psi_for_negation}
\end{aligned}
\end{equation}
Substituting Eq. \eqref{eq_EV_relationship_between_Psi_prime_and_Psi_for_negation} for $i=3$ into the left hand side of Eq. \eqref{eq_EV_relationship_between_Psi_prime_and_Psi_for_phi1_andphi3} proves Eq. \eqref{eq_theorem_1_part_1}.  Substituting Eq. \eqref{eq_EV_relationship_between_Psi_prime_and_Psi_for_negation} for $i = 2$ into the left hand side of Eq. \eqref{eq_EV_relationship_between_Psi_prime_and_Psi_for_phi2} proves Eq. \eqref{eq_theorem_1_part_2}.  Q.E.D.

The following corollary follows trivially from Theorem 1.

\noindent \underline{\textbf{Corollary 1}}

Under the same assumptions and definitions stated in Theorem 1, for all times $t$,
\begin{equation}
\langle A_{even}(\phi_1) \rangle(t) = \langle A_{even}(\phi_3) \rangle(t), \label{eq_Corollary_1_result_1}
\end{equation}
\begin{equation}
\langle A_{odd}(\phi_1) \rangle(t) = -\langle A_{odd}(\phi_3) \rangle(t), \label{eq_Corollary_1_result_2}
\end{equation}
and
\begin{equation}
\langle A_{odd}(\phi_2) \rangle(t) = 0, \label{eq_Corollary_1_result_3}
\end{equation}
where 
$A_{even}(\phi_i)$ is any arbitrary observable of a single degree of freedom that satisfies $A_{even}(\phi_i) = A_{even}(-\phi_i)$, and $A_{odd}(\phi_i)$ is any arbitrary observable of a single degree of freedom that satisfies $A_{odd}(\phi_i) = -A_{odd}(-\phi_i)$.

\noindent \underline{\textbf{Theorem 2}}

Consider the Hamiltonian $H$ defined in Section \ref{Subsubsection_Model_Hamiltonian_for_Three_Coupled_Molecular_Rotors} with $R_{12} = R_{13} = R_{23} = R$ and $\theta_{12} = \frac{\pi}{3}$, $\theta_{13} = 0$, and $\theta_{23} = \frac{5 \pi}{3}$, and initial state $\ket{\Psi_0} = \ket{0} \otimes \ket{0} \otimes \ket{0}$.  For any arbitrary field $\varepsilon(t)$ (not necessarily optimal), Eqs.~\eqref{eq_cosphi1_cosphi_3_relationship}, \eqref{eq_sinphi1_sinphi3_relationship}, and \eqref{eq_sinphi2_relationship} hold at all times $t$.

\noindent \underline{\textbf{Proof of Theorem 2}}

First, we show that $H$ satisfies Eq. \eqref{eq_condition_1_of_theorem_1}.  For $H^{(0)}$, it is trivial that
\begin{equation}
\begin{aligned}
H^{(0)}(\phi_1,\phi_2,\phi_3;t) &= H^{(0)}(\phi_1 \rightarrow \phi_3,\phi_2,\phi_3 \rightarrow \phi_1;t) \\
&= H^{(0)}(-\phi_1,-\phi_2,-\phi_3;t).
\end{aligned}
\end{equation}
Therefore, it suffices to show that
\begin{equation}
W(\phi_1 \rightarrow \phi_3, \phi_2, \phi_3 \rightarrow \phi_1) = W(-\phi_1, -\phi_2, -\phi_3) \label{eq_W_perm_equals_W_fsc}
\end{equation}
Substituting $R_{12} = R_{13} = R_{23} = R$ and $\theta_{12} = \frac{\pi}{3}$, $\theta_{13} = 0$, and $\theta_{23} = \frac{5 \pi}{3}$ into $W$ and rearranging, we have:
\begin{equation}
W(\phi_1,\phi_2,\phi_3) = W_{sym}(\phi_1,\phi_2,\phi_3) + W_{anti-sym}(\phi_1,\phi_2,\phi_3),
\end{equation}
where
\begin{equation}
\begin{aligned}
W&_{sym}(\phi_1,\phi_2,\phi_3) = \frac{\mu^2}{4 \pi \epsilon_0 R^3} \Big( \frac{1}{4} \cos\phi_2(\cos\phi_1 + \cos\phi_3)\\
&- \frac{5}{4}\sin\phi_2(\sin\phi_1 + \sin\phi_3) - 2\cos\phi_1 \cos\phi_3 + \sin\phi_1 \sin\phi_3\Big),
\end{aligned}
\end{equation}
and 
\begin{equation}
\begin{aligned}
W_{anti-sym} &= -\frac{3 \sqrt[]{3}}{4} \Big( \cos\phi_2(\sin\phi_1 - \sin\phi_3) \\
&+ \sin\phi_2 (\cos\phi_1 - \cos\phi_3) \Big).
\end{aligned}
\end{equation}
As such, we can see that
\begin{equation}
\begin{aligned}
W_{sym}(\phi_1 \rightarrow \phi_3,\phi_2,\phi_3 \rightarrow \phi_1) &= W_{sym}(\phi_1,\phi_2,\phi_3) \\
&= W_{sym}(-\phi_1,-\phi_2,-\phi_3)
\end{aligned}
\end{equation}
and
\begin{equation}
\begin{aligned}
W_{anti-sym}(&\phi_1 \rightarrow \phi_3,\phi_2,\phi_3 \rightarrow \phi_1) \\
&= - W_{anti-sym}(\phi_1,\phi_2,\phi_3) \\
&= W_{anti-sym}(-\phi_1,-\phi_2,-\phi_3).
\end{aligned}
\end{equation}
Eq. \eqref{eq_W_perm_equals_W_fsc} follows, and so Eq. \eqref{eq_condition_1_of_theorem_1} is satisfied.  Additionally, $\ket{\Psi_0}$ satisfies Eq. \eqref{eq_condition_2_of_theorem_1}, which can be seen by the fact that in coordinate space, $\ket{0} \otimes \ket{0} \otimes \ket{0} = \frac{1}{(2\pi)^{3/2}}$.  Therefore, we can apply Corollary 1, and Eqs. \eqref{eq_cosphi1_cosphi_3_relationship}, \eqref{eq_sinphi1_sinphi3_relationship}, and \eqref{eq_sinphi2_relationship} follow as specific cases of Eqs. \eqref{eq_Corollary_1_result_1}, \eqref{eq_Corollary_1_result_2}, and \eqref{eq_Corollary_1_result_3} respectively.  Q.E.D.

\end{document}